\documentclass[10pt,twocolumn,letterpaper]{article}

\usepackage[pagenumbers]{cvpr}

\definecolor{cvprblue}{rgb}{0.21,0.49,0.74}
\usepackage[pagebackref,breaklinks,colorlinks,allcolors=cvprblue]{hyperref}
\usepackage{multirow}
\usepackage{amssymb}
\usepackage{pifont}
\usepackage{makecell}
\usepackage{wrapfig}
\usepackage{booktabs}
\usepackage{multirow}
\usepackage{subcaption}
\usepackage{graphicx}
\usepackage{booktabs}
\usepackage{multirow}

\newcommand{\dbname}{STAM\xspace}
\newcommand{\name}{MoSAT\xspace}


\title{\name: Human Motion Generation from Spatial Audio and Textual Description}
\author{Shuyang Xu$^{*,1}$, Zhiyang Dou$^{\dag,1,2}$, Yiduo Hao$^{3}$, Zekun Li$^{4}$, Liang Pan$^{1}$, \\
Jingbo Wang$^{5}$, Cheng Lin$^{6}$, Yuan Liu$^{7}$, Wenping Wang$^{\dag,8}$, Mingmin Zhao$^{3}$, Taku Komura$^{\dag,1}$
\\
\\
$^{1}$The University of Hong Kong \quad
$^{2}$Massachusetts Institute of Technology \\
$^{3}$University of Pennsylvania \quad
$^{4}$Brown University \quad
$^{5}$Shanghai AI Lab \\
$^{6}$Macau University of Science and Technology \\
$^{7}$Hong Kong University of Science and Technology \\
$^{8}$Texas A\&M University\\
}

\begin{document}
\maketitle
\begin{abstract}
Human motion is shaped by both external acoustic events and behavioral intent: spatial audio conveys environmental cues that elicit or guide a response, while text specifies the desired action and how it should be performed. In this paper, we study the novel task of human motion synthesis jointly conditioned on spatial audio and natural language, a problem that has been largely overlooked in previous research. To support this task, We introduce \textbf{\dbname}, a dataset of motion sequences paired with spatial audio and detailed textual annotations whose rich vocabulary affords precise and nuanced specification of human motions. We further introduce \textbf{\name}, a latent flow-matching framework for full-body motion generation jointly conditioned on natural-language intent and directional spatial-audio cues through hierarchical cross-attention before generating motion. Such a hierarchical design enhances temporally coherent and semantically aligned motion sequences. We also develop tri-modal evaluators for comprehensive evaluation on this novel task. Extensive experiments show that \name achieves the SOTA performance by leveraging spatial audio’s intrinsic motion-shaping properties alongside textual semantics, enabling precise and diverse motion in various scenarios.
\end{abstract}
\vspace{-4mm}    
\section{Introduction}
\label{sec:intro}

Human motion is shaped by both what a person intends to do and what unfolds around them. 
Language can specify an action and its style (e.g., by using the language prompt \emph{turn toward the ringing phone and reach cautiously}), but encoding the exact onset, duration, and source direction of the event at every moment in language is cumbersome.
Audio provides temporally dense acoustic evidence, while an associated DOA anchor grounds the event relative to the body.
The two modalities therefore play complementary roles: language specifies \emph{what} motion to perform and \emph{how} to perform it, whereas spatial and acoustic cues anchors \emph{where} and \emph{when} the motion should unfold.

Existing motion generators, however, largely study these signals separately. 
Text-to-motion methods provide semantic control over actions and styles~\cite{guo2024momask, zhou2024emdm, tevet2022human, chen2023executing, zhang2024motiondiffuse, petrovich2024multi, petrovich2022temos, zhang2023generating}, 
while audio-driven methods use speech or music to govern gesture, rhythm, and timing~\cite{zhang2024semantic, cheng2024siggesture, zhu2023taming, chen2025language, li2022danceformer, tseng2023edge, gong2023tm2d, dabral2023mofusion}. 
Most recently, MoSPA~\cite{xu2026mospa} introduces spatial audio as a motion condition, grounding synthesis in directional acoustic cues, but it does not provide language-based control over semantic intent. 
This separation prevents users from jointly specifying a desired behavior and binding it to the geometry and temporal evolution of an acoustic event. 
We therefore introduce the task of human motion generation jointly conditioned on natural language and spatial audio, allowing users to express high-level intent through language while spatial audio governs the resulting motion's precise trajectory and timing. The central obstacle is the absence of paired training data, an evaluation protocol, and strong cross-modal conditioning for this task. 

We therefore curate \textbf{\dbname}, pairing motion sequences with time-synchronized binaural audio and motion-aligned textual captions spanning diverse actions and styles. The corpus contains over \emph{194K} words with a vocabulary of \emph{1,783}, supporting both diversity and precision of annotation
\footnote{We assume that linguistic intent and spatial audio cues are consistent--a standard assumption in multimodal-to-unimodal generation tasks.}, which enables fine-grained specification of spatial-audio–conditioned behaviors and styles, and establishes a unified testbed for this task.
Building on \dbname, we study human motion generation jointly conditioned on spatial audio and text. The key challenge is aligning these heterogeneous modalities—continuous acoustic dynamics, discrete linguistic semantics, and high-dimensional motion—to generate realistic, spatially grounded, and semantically faithful motion.

To address this challenge, we present \textbf{\name}, a novel framework designed to coordinate textual and spatial audio context for motion synthesis while establishing a standardized baseline for benchmarking. Architecturally, a Motion Variational Autoencoder (\emph{VAE}) maps complex skeletal motion into a compact, expressive latent space, while the pre-trained CLAP~\cite{elizalde2023clap} encoder extracts frame-aligned acoustic semantic features. To incorporate spatial geometry, we explicitly condition the framework on the normalized Direction of Arrival (DOA) of the sound event computed at the initial frame, establishing a robust spatial anchor for audio-oriented motion trajectories.

Conditioned on these unified multimodal representations, \textbf{\name} leverages a multi-head cross-attention mechanism to fuse linguistic intent with spatial and acoustic cues. A transformer-based flow matching model then generates continuous trajectories within the motion latent space. This generative formulation enables flexible and controllable trade-offs between adherence to directional audio signals and high-level text specifications, all while maintaining high perceptual quality and kinematic realism.

To fill the current gap in evaluating text- and spatial-audio-conditioned motion generation, we design dedicated tri-modal retrieval protocols inspired by prior text--motion retrieval literature~\cite{guo2022generating, petrovich2023tmr, bensabath2024cross}. Specifically, we introduce Condition--Motion Retrieval (\textbf{CMR}), which tests whether multimodal condition queries retrieve the corresponding motion, and Motion--Condition Retrieval (\textbf{MCR}), which evaluates whether a motion clip retrieves its paired condition---where a condition incorporates the textual description, audio features, and Direction of Arrival (DOA). Extensive experiments show that \name achieves SOTA alignment and fidelity by exploiting spatially positioned audio's motion-shaping cues alongside textual semantics, yielding precise and diverse motion across multiple scenarios and datasets. These quantitative protocols, complementary qualitative analyses, and ablation studies probing condition selection and model size jointly validate our key design choices. In summary, our primary contributions are:
\begin{itemize}
    \item We introduce \textbf{\dbname}, a dataset that pairs motion with spatial audio and diverse, motion-aligned textual annotations, establishing a unified benchmark that stresses both spatial grounding and semantic control.
    \item We propose \name, a baseline architecture for text- and spatial-audio-conditioned motion synthesis, combining a Motion VAE with a Transformer-based flow matching model to leverage complementary semantic and spatial cues.
    \item We develop tri-modal evaluators--\textbf{CMR} and \textbf{MCR}--built with a contrastively trained scorer for evaluation. \name demonstrates the SOTA performance against baselines through extensive experiments.
\end{itemize}
\section{Related Works}
\label{sec:related}

\noindent
\textbf{Text-to-Motion.}
Methods conditioned on natural language descriptions~\cite{athanasiou2022teach, athanasiou2023sinc, chen2024pay, tevet2022human, zhou2024emdm, chen2023executing, zhang2024motiondiffuse, petrovich2024multi, petrovich2022temos} or action-level signals~\cite{guo2020action2motion, xu2023actformer, petrovich2021action} have been extensively explored in recent years. Diffusion-based methods, including MDM~\cite{tevet2022human}, EMDM~\cite{zhou2024emdm}, MLD~\cite{chen2023executing}, and MotionDiffusion~\cite{zhang2024motiondiffuse}, as well as MoMask~\cite{guo2024momask}, which employs discrete masked generative modeling over quantized motion representations, achieve high-quality motion generation with strong semantic alignment. Another line of work~\cite{lu2025scamo, fan2025go, zhang2023generating} scales up motion generation using autoregressive frameworks, where motions are first quantized~\cite{mentzer2024finite, van2017neural, lee2022autoregressive} and then generated token-by-token. While language signals are powerful for specifying high-level intent, they fall short in conveying fine-grained cues that govern detailed body dynamics. In this work, we address this limitation by learning a generative framework jointly conditioned on textual descriptions and spatial audio. Unlike existing approaches, our architecture integrates a motion VAE with a Transformer-based flow matching model, leveraging tailored cross-attention mechanisms to effectively capture complex multimodal dependencies. Furthermore, our formulation enables flexible inference-time editing, offering enhanced control for downstream applications.

\noindent
\textbf{Audio-to-Motion.}
A large body of work studies motion synthesis conditioned on audio, spanning speech~\cite{zhang2024semantic, cheng2024siggesture, ao2023gesturediffuclip, ao2022rhythmic, zhu2023taming, chen2025language} and music~\cite{siyao2022bailando, li2022danceformer, tseng2023edge, gong2023tm2d, yang2025unimumo, dabral2023mofusion}. 
In speech-driven motion synthesis, earlier work predominantly employed GAN-based models~\cite{ginosar2019learning, liu2022learning, qian2021speech, yoon2020speech}. More recent approaches have transitioned toward diffusion frameworks~\cite{zhu2023taming}. Rhythmic gesticulator~\cite{ao2022rhythmic} improves temporal correspondence by segmenting speech and disentangling speech–motion representations. Semantic Gesticulator~\cite{zhang2024semantic} proposes retrieving gestures that are semantically aligned with speech using an LLM-enhanced gesture library. Furthermore, Chen et al.~\cite{chen2025language} unifies speech, text, and motion within a single multimodal autoregressive transformer.
For music-conditioned motion synthesis, DanceFormer~\cite{li2022danceformer} generates beat-aligned key poses and subsequently refines them into continuous motion trajectories. 

Alexanderson et al.~\cite{alexanderson2023listen} employ a Conformer-based diffusion framework for high-fidelity motion synthesis, while EDGE~\cite{tseng2023edge} extends diffusion models to controllable dance generation and editing. Multimodal approaches jointly modeling language and audio further improve semantic coherence and user controllability~\cite{gong2023tm2d, yang2025unimumo, dabral2023mofusion, bian2025motioncraft, zhang2025motion}. MoSpa~\cite{xu2026mospa} introduces spatial audio as a conditioning signal for motion synthesis, but spatial acoustics have not been jointly explored with natural language for semantic-aware motion synthesis. To address this gap, we curate the dataset in~\cite{xu2026mospa} containing synchronized motion, spatial audio, and diverse motion-aligned textual descriptions, and introduce a method for high-fidelity spatial-audio–driven motion synthesis with rich semantic control through text.

\noindent
\textbf{Flow-based Motion Generation.}
Flow matching has recently emerged as an alternative to diffusion-based motion generation, enabling continuous transport modeling with efficient sampling~\cite{hu2023motion, li2026motionhiflow, gupta2026unified}. We build upon this direction by applying flow matching in a learned motion latent space and jointly incorporating text, acoustic semantics, and spatial conditions.
\section{Method}
\label{sec:method}

\begin{figure*}[!t]
    \centering
    \vspace{-3mm}
    \includegraphics[width=0.9\linewidth]{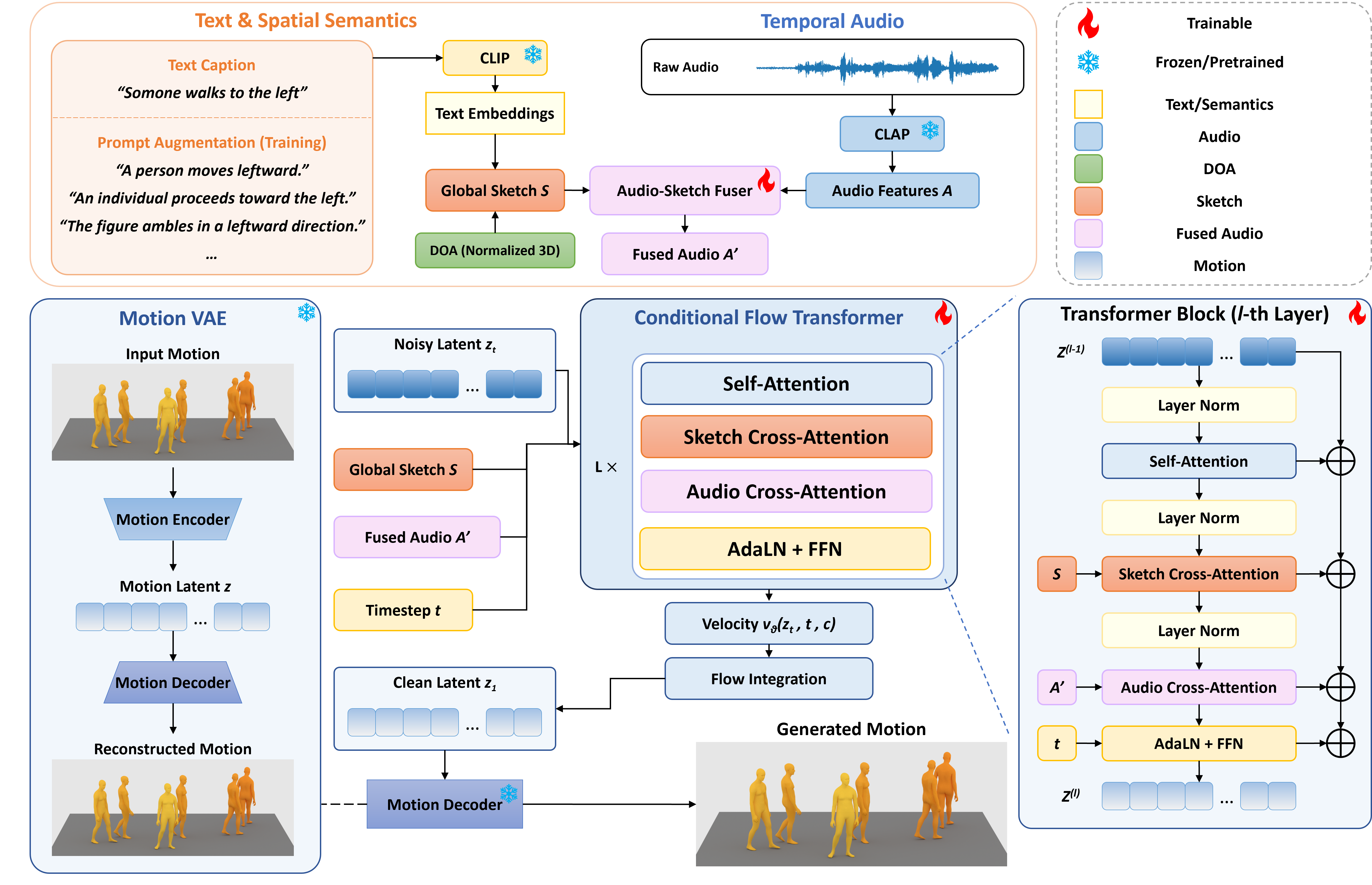}
    \caption{
    \textbf{(Left)} A Motion VAE is pre-trained to compress high-dimensional ($272\text{D}$) motion features into a compact and expressive latent space. 
    \textbf{(Top)} Heterogeneous multimodal inputs, including text captions, spatial audio, and direction-of-arrival (DOA) information, are fused through a multi-head cross-attention mechanism within the Audio-Sketch Fuser. 
    \textbf{(Bottom Right)} The resulting multimodal context conditions a Transformer-based flow matching model together with noisy motion latent tokens to predict vector-field velocities, from which high-fidelity motion is synthesized. All sub-blocks in the flow matching transformer are implemented with pre-normalization and residual connections.
    }
    \label{fig:pipeline}
    \vspace{-5mm}
\end{figure*}

As illustrated in Fig.~\ref{fig:pipeline}, our framework consists of two primary components. First, a Motion Variational Autoencoder (VAE) compresses continuous, high-dimensional skeletal motion into a compact and expressive latent representation. Second, a Transformer-based flow matching model operates in this latent space to synthesize motion conditioned jointly on text, spatial audio, and direction-of-arrival (DOA) information. To extend the model beyond the sequence lengths used during training, we further adopt a masked training and inference strategy similar to EDGE~\cite{tseng2023edge}. In addition, we introduce an inference-time joint-level editing mechanism that provides fine-grained control over the synthesized motion while preserving multimodal condition alignment and kinematic fidelity.

\subsection{Dataset}

We use the \dbname~dataset to train our proposed model, \name. To enrich the multimodal supervision, we extend the SAM~\cite{xu2026mospa} dataset by adding comprehensive textual descriptions for its motion sequences, resulting in the creation of \dbname. Specifically, starting from the complete SAM release, we remove 20 bad sequences, retaining 3,340 valid ten-second clips, and each motion is annotated with 1 manual caption and three human verified DeepSeek-assisted~\cite{liu2024deepseek} variants. Among these descriptions, some explicitly mention the sound type and the spatial location of the sound source, while others describe the motion only. In total, \dbname~contains over 194K words, spanning 1,783 distinct vocabulary terms. This diversity of textual annotations enhances the richness of the dataset and strengthens the generalization capability of \name.

\subsection{Motion VAE}

\noindent\textbf{Motion Representation.} 
Following~\cite{xiao2025motionstreamer}, we represent human motion at each frame $t$ as a 272-dimensional vector $\mathbf{x}_t = [\mathbf{v}_{\text{root}}, \boldsymbol{\omega}_{\text{root}}, \mathbf{p}, \mathbf{v}, \mathbf{r}] \in \mathbb{R}^{272}$. We select $J = 22$ body joints corresponding to the primary kinematic tree of the SMPL/SMPL-X models~\cite{loper2015smpl, pavlakos2019expressive}. The sub-vector components are defined as follows:
\begin{itemize}
    \item $\mathbf{v}_{\text{root}} \in \mathbb{R}^2$: Heading-aligned root linear velocity on the $XZ$ ground plane.
    \item $\boldsymbol{\omega}_{\text{root}} \in \mathbb{R}^6$: Continuous 6D root heading rotation.
    \item $\mathbf{p} \in \mathbb{R}^{3J}$: Heading-stripped local 3D joint positions.
    \item $\mathbf{v} \in \mathbb{R}^{3J}$: Heading-stripped local linear joint velocities.
    \item $\mathbf{r} \in \mathbb{R}^{6J}$: Local 6D joint rotations.
\end{itemize}
All orientation and rotation terms utilize the 6D continuous representation~\cite{zhou2019continuity} to prevent rotational discontinuities. Unlike the standard 263-dimensional HumanML3D feature space~\cite{guo2022generating}, this formulation eliminates explicit binary foot-contact channels while implicitly embedding root height within the position block $\mathbf{p}$. The global ground trajectory is recovered via numerical integration of $\mathbf{v}_{\text{root}}$. Prior to model training, all features are per-dimension $z$-normalized using dataset mean and standard deviation to match a standard normal distribution. Further details of the features are provided in Appendix~\ref{app:motion_features}.

\noindent\textbf{Architecture and Training Objectives.} 
To compress motion sequences into an expressive latent manifold, we employ a 1D convolutional VAE with residual blocks, inspired by~\cite{guo2024momask, xiao2025motionstreamer}. The encoder downsamples temporal sequences by a factor of $4\times$ into a latent dimension of $16$. The model is trained on sliding windows of $64$ frames extracted from HumanML3D and~\dbname.

The Motion VAE is optimized end-to-end via a composite loss function:
\begin{equation}
    \mathcal{L}_{\text{vae}} = \mathcal{L}_{\text{recon}} + w_{\text{fk}} \mathcal{L}_{\text{fk}} + w_{\text{root}} \mathcal{L}_{\text{root}} + w_{\text{KL}} \mathcal{L}_{\text{KL}}
\end{equation}
where $\mathcal{L}_{\text{recon}} = \|\mathbf{x} - \mathbf{x}'\|_{\text{smooth\_L1}}$ measures the Smooth L1 reconstruction error between input motion $\mathbf{x}$ and reconstructed motion $\mathbf{x}'$. The term $\mathcal{L}_{\text{fk}} = \|\text{FK}(\mathbf{x}) - \text{FK}(\mathbf{x}')\|_1$ enforces kinematic consistency, where $\text{FK}(\cdot)$ denotes forward kinematics computing global 3D joint positions from 6D rotations~\cite{zhou2019continuity}. $\mathcal{L}_{\text{root}}$ provides targeted supervision on the root trajectory to mitigate drift, and $\mathcal{L}_{\text{KL}}$ penalizes latent distribution divergence with a weight $w_{\text{KL}}$ linearly warmed up over the first 15 epochs.

\subsection{Flow Matching Motion Generation}

\noindent\textbf{Problem Formulation and Flow Matching Preliminaries.}
Given a compact motion latent representation $\mathbf{z}_1 \in \mathbb{R}^{T_z \times d_m}$ extracted by the Motion VAE from a sequence of $T_f=300$ motion frames, where $T_z=75$ denotes the number of latent tokens under a temporal downsampling factor of $4\times$, our objective is to model the conditional distribution $p(\mathbf{z}_1 \mid \mathbf{c})$. Here, $\mathbf{c}=(\mathbf{T},\mathbf{A},\mathbf{d})$ denotes the multimodal conditioning tuple, consisting of a textual description $\mathbf{T}$, frame-aligned acoustic features $\mathbf{A}$, and the direction of arrival (DOA) of the sound source $\mathbf{d}$.

We formulate conditional motion generation using Continuous Normalizing Flows through Conditional Flow Matching (CFM)~\cite{lipman2022flow}. Rather than explicitly simulating a diffusion process, CFM learns a time-dependent conditional vector field $\mathbf{v}_\theta(\mathbf{z}_t,t,\mathbf{c})$ that transports samples from a simple Gaussian prior $\mathbf{z}_0 \sim \mathcal{N}(\mathbf{0},\mathbf{I})$ to the target motion distribution. Specifically, assuming a linear interpolation between the noise and data distributions, the intermediate state $\mathbf{z}_t$ at continuous time $t\in[0,1]$ and its corresponding target velocity $\mathbf{u}_t$ are defined as
\begin{equation}
    \mathbf{z}_t = t\mathbf{z}_1 + (1-t)\mathbf{z}_0,
    \qquad
    \mathbf{u}_t(\mathbf{z}_t) = \frac{d\mathbf{z}_t}{dt}
    = \mathbf{z}_1 - \mathbf{z}_0.
\end{equation}

\vspace{-2mm}
\noindent\textbf{Multimodal Conditioning Streams.}
To process heterogeneous inputs with distinct spatial and temporal characteristics, we organize the conditioning information into three input streams, each projected into the Transformer latent dimension $d_{\text{model}}=512$:
\begin{enumerate}
    \item \textbf{Global Sketch Stream ($\mathbf{S} \in \mathbb{R}^{L_s \times d_{\text{model}}}$):}
    Captures high-level semantic and spatial context using $L_s=33$ token positions. The stream is constructed by concatenating the linear projection of the initial sound source direction $\mathbf{d}$, normalized as the direction of arrival (DOA) relative to the subject, with the global CLIP text embedding~\cite{radford2021learning}, the first $30$ fine-grained word-level CLIP tokens, and a neutral genre token. Following~\cite{xu2026mospa}, we retain but neutralize the genre-token slot, as text already captures motion intensity while coarse genre cues may introduce ambiguous or conflicting signals. See details in Sec.~\ref{sec:exp}.
    
    \item \textbf{Temporal Acoustic Stream ($\mathbf{A} \in \mathbb{R}^{T_z \times d_{\text{model}}}$):}
    Encodes frame-aligned acoustic information extracted from CLAP embeddings~\cite{elizalde2023clap}. The acoustic features are extracted at $7.5\text{ Hz}$, resulting in a sequence length that matches the temporal length $T_z=75$ of the compressed motion representation.
    
    \item \textbf{Noisy Motion:}
    Represents the noisy motion state at continuous timestep $t$, obtained by linearly projecting the interpolated motion latent $\mathbf{z}_t$ into the Transformer latent space.
\end{enumerate}

\noindent\textbf{Architecture Details.}
As detailed in Fig.~\ref{fig:pipeline}, the flow matching backbone consists of an Audio-Sketch Fusion Encoder followed by a stack of $L$ conditional Transformer blocks. The noisy motion trajectory is augmented with a per-frame binary hole mask, where $1$ indicates frames to be generated and $0$ indicates anchored frames. This mask provides an explicit mechanism for conditioning on known motion and enables EDGE-style~\cite{tseng2023edge} chunked generation of long sequences.

\textit{1) Audio-Sketch Fusion Encoder:}
Prior to interacting with the motion tokens, the acoustic sequence $\mathbf{A}$ attends to the global sketch sequence $\mathbf{S}$ through cross-attention, producing grounded acoustic tokens $\mathbf{A}'$:
\begin{equation}
    \mathbf{A}' =
    \operatorname{Softmax}\left(
    \frac{\mathbf{A}\mathbf{W}_Q(\mathbf{S}\mathbf{W}_K)^\top}
    {\sqrt{d_{\text{head}}}}
    \right)
    \mathbf{S}\mathbf{W}_V + \mathbf{A},
\end{equation}
where $\mathbf{W}_Q$, $\mathbf{W}_K$, and $\mathbf{W}_V$ denote learned projection matrices. The sketch sequence $\mathbf{S}$ contains a neutralized genre token, a spatial direction-of-arrival (DOA) token, a sentence-level text embedding, and fine-grained word-level contextual embeddings. This early fusion step grounds temporal acoustic variations in the global textual and spatial context before they are integrated with the motion representation.

\textit{2) Conditional Motion Transformer:}
Each Transformer block sequentially processes the motion tokens using self-attention, sketch cross-attention, audio cross-attention, and a timestep-conditioned AdaLN-FFN, with residual connections at each stage. The sketch stream provides global textual and spatial context, while the acoustic stream injects temporally aligned audio information. The continuous flow timestep $t$ modulates the AdaLN-FFN, as illustrated in Fig.~\ref{fig:pipeline}.

\noindent\textbf{Training Objective.}
The flow matching model is trained to predict the vector field $\mathbf{v}_\theta(\mathbf{z}_t,t,\mathbf{c})$ using a masked mean squared velocity regression loss. Let $\mathbf{m}\in\{0,1\}^{T_z}$ denote a binary motion mask, where valid motion tokens are assigned $1$ and padded tokens are assigned $0$. The training objective is
\begin{equation}
    \mathcal{L}_{\text{FM}}(\theta)
    =
    \mathbb{E}_{t,\mathbf{z}_0,\mathbf{z}_1}
    \left[
    \mathbf{m}\odot
    \left\|
    \mathbf{v}_\theta(\mathbf{z}_t,t,\mathbf{c})
    -(\mathbf{z}_1-\mathbf{z}_0)
    \right\|_2^2
    \right]
\end{equation}
where continuous timesteps are sampled from
$t\sim\operatorname{LogitNormal}(0,1)$.

\begin{figure}[htbp]
    \vspace{-2mm}
    \centering
    \includegraphics[width=\linewidth]{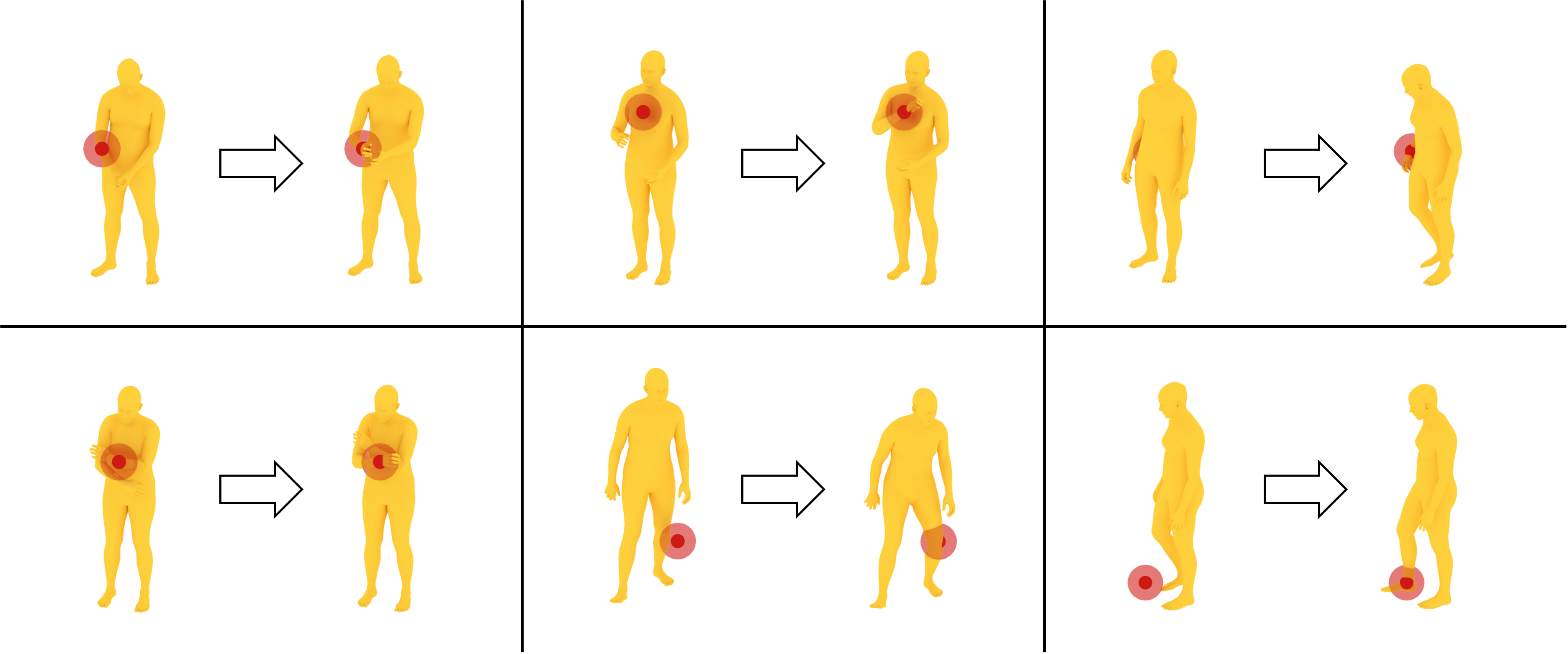}
    \caption{Visualization of joint constraint editing. A joint constraint is specified at a particular frame, with red spheres indicating the target joint positions. The synthesized motion accurately satisfies the prescribed constraints while maintaining smooth and natural motion.}
    \label{fig:edit}
    \vspace{-6mm}
\end{figure}

\subsection{Inference and Editing}

A key advantage of our flow matching formulation is its support for zero-shot motion editing at inference time without retraining or fine-tuning.

\noindent\textbf{Keyframe Joint Control.} To satisfy sparse spatial constraints, such as target 3D joint positions at selected frames, we apply gradient-based latent guidance inspired by Diffusion Posterior Sampling (DPS)~\cite{chung2022diffusion} and optimization-based motion guidance methods~\cite{karunratanakul2023guided, karunratanakul2024optimizing}. After each integration step, the latent $\mathbf{z}_t$ is decoded by the frozen VAE, $\hat{\mathbf{x}}_t=\text{Decoder}(\mathbf{z}_t)$, and an editing loss
\begin{equation}
    \mathcal{L}_{\text{edit}}
    = \left\|\text{FK}(\mathbf{x})_{\text{pred}}-\text{FK}(\mathbf{x})_{\text{target}}\right\|_2^2
\end{equation}
is computed at the constrained frames. We backpropagate this loss through the frozen decoder and update the latent as
\begin{equation}
    \mathbf{z}_t \leftarrow
    \mathbf{z}_t-\lambda\nabla_{\mathbf{z}_t}\mathcal{L}_{\text{edit}},
\end{equation}
repeating this decode--optimize cycle for a few inner iterations before continuing integration. Root-trajectory editing follows the same procedure, constraining the root's horizontal-velocity channels (whose cumulative sum traces the root trajectory) over a window of 
frames. Because guidance operates in the learned motion latent space, the resulting edits remain close to the pretrained motion manifold. See Fig.~\ref{fig:edit} for joint constraint editing examples.

\noindent\textbf{Long-Sequence Generation.}
Following EDGE~\cite{tseng2023edge}, we train the model with randomly masked contiguous motion regions and provide the corresponding binary hole mask as an additional input channel. At inference, overlapping chunks are generated sequentially, with each new chunk anchored to the tail of the preceding output. The anchored frames are constrained to follow the interpolated path toward the preceding clean motion, while the remaining frames are generated afresh.

\noindent\textbf{Masking and Classifier-Free Guidance}
We employ modality dropout and Classifier-Free Guidance (CFG)~\cite{ho2022classifier, guo2024momask} during training and inference. For samples containing all modalities, text, audio, and spatial DOA are independently dropped with probability $0.15$, while all conditions are jointly dropped with probability $0.2$. The genre token is always neutralized. Separately, contiguous motion regions are masked with probability $0.5$ to train the known-motion conditioning described above. At inference, CFG is applied against null conditions (neutral genre, null DOA, empty text, and zero audio) to improve motion quality and controllability. We find that the CFG scale affects different evaluation metrics differently; further analysis is provided in Appendix~\ref{app:experiments}.
\section{Experiments}
\label{sec:exp}

\subsection{Implementation Details}

\begin{figure}[htbp]
    \vspace{-2mm}
    \centering
    \includegraphics[width=0.9\linewidth]{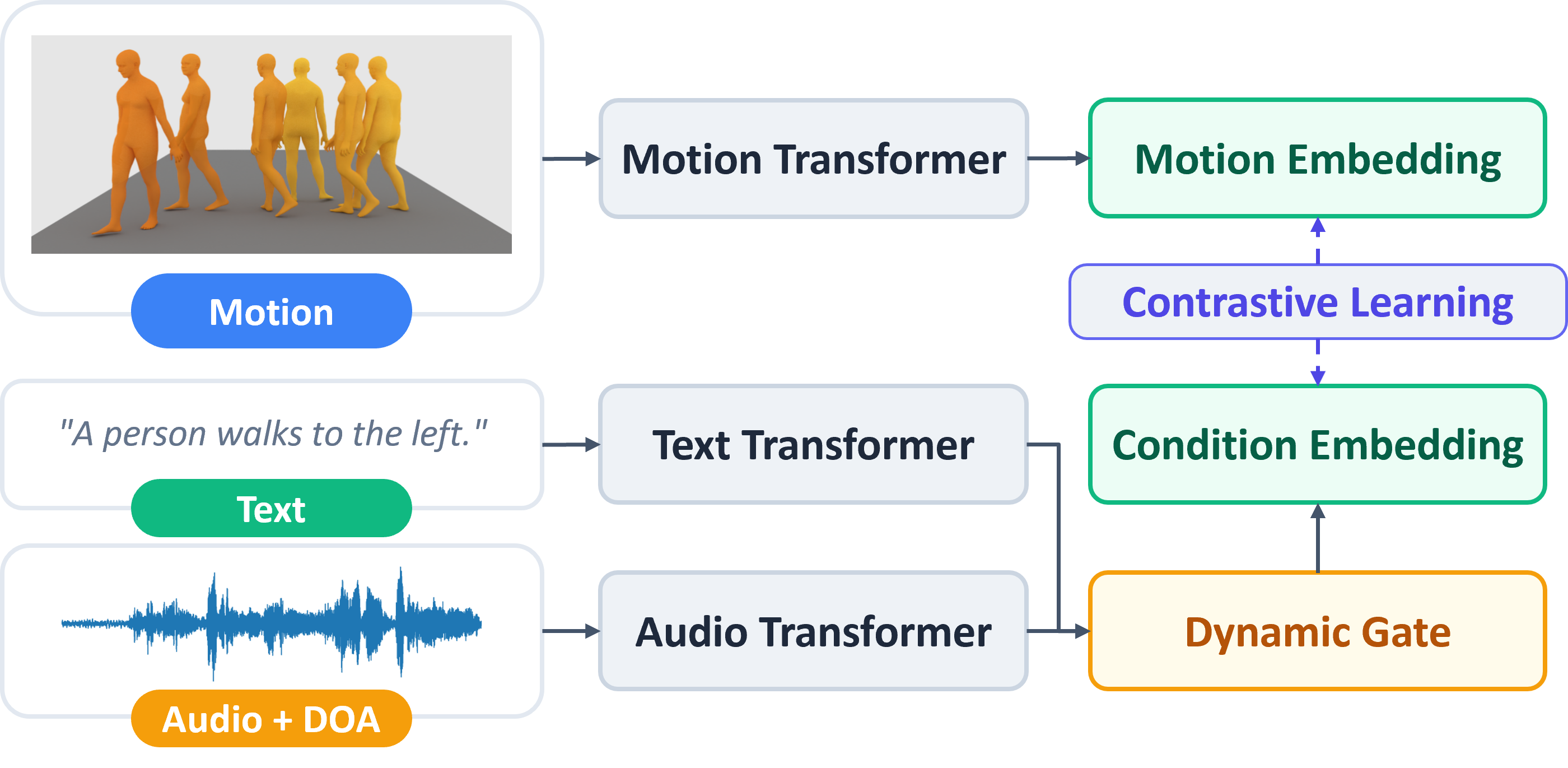}
    \caption{Overview of the proposed tri-modal evaluator. Motion, text, and spatial-audio inputs are encoded into a shared embedding space. The text and spatial-audio embeddings are dynamically fused to form a unified multimodal condition representation, which is trained to align with the corresponding motion embedding using a symmetric contrastive objective.
    }
    \label{fig:evaluator_main}
    \vspace{-4mm}
\end{figure}

\begin{figure*}
    \centering
    \vspace{-1mm}
    \includegraphics[width=0.85\linewidth]{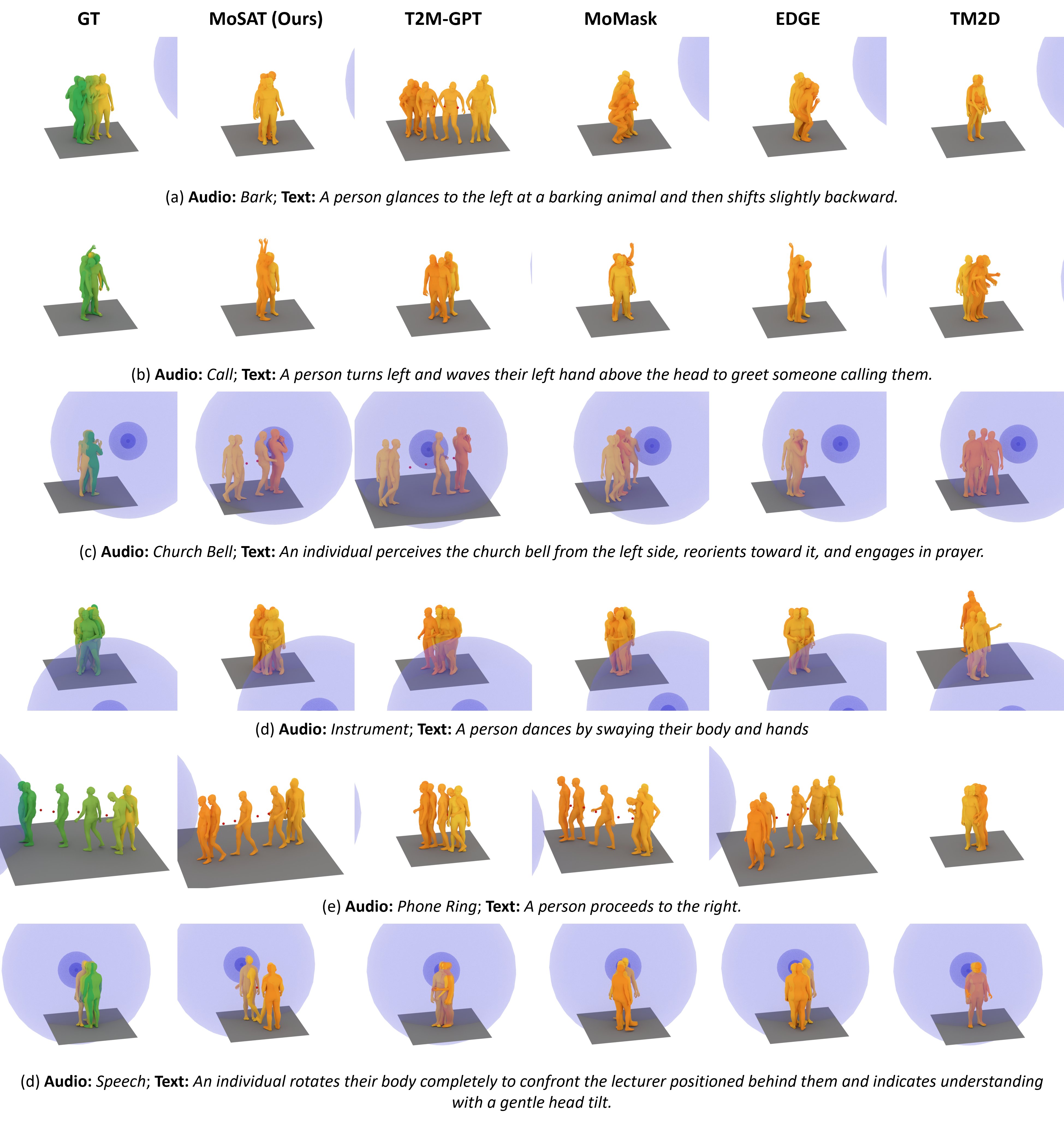}
    \vspace{-3mm}
    \caption{Qualitative comparison between \name and baseline models, where \name yields more realistic and responsive human motion compared to baseline models. Here, the red dots show the character's trajectory. The blue semitransparent sphere represents the approximate location of the sound source.
    }
    \label{fig:qualitative}
    \vspace{-5mm}
\end{figure*}

\noindent\textbf{Joint Multimodal Co-Training.}
To improve text-conditioned controllability and leverage large-scale motion priors, we jointly train our model on HumanML3D~\cite{guo2022generating} and our \dbname dataset. Since HumanML3D lacks acoustic annotations, we set its audio features to zero and use a default DOA vector $\mathbf{d}=(0,1,0)$.

\noindent\textbf{Data Augmentation.}
For \dbname, we apply spatial mirroring and text dropping to improve spatial generalization and multimodal responsiveness. Specifically, mirrored motions are generated by reflecting skeletal trajectories along the $Z$-axis in SMPL space, with the corresponding DOA vectors spatially reflected while preserving the acoustic features. The dataset is divided into training, validation, and test sets with a ratio of 0.8:0.05:0.15, with mirrored samples carefully accounted for during the splitting process. We additionally create a text-free copy of \dbname, allowing the model to learn motion generation from spatial audio and DOA alone.

\noindent\textbf{Training Details.} All experiments are performed on a single NVIDIA RTX~4090 GPU. The Motion VAE is trained with AdamW~\cite{loshchilov2017decoupled} using a batch size of 256 and an initial learning rate of $2\times10^{-4}$, with a MultiStep scheduler ($\gamma=0.05$) at iterations $[150000,250000]$. The flow matching model is trained for 500 epochs with AdamW, a batch size of 64, and an initial learning rate of $2\times10^{-4}$ using a cosine scheduler decaying to zero. Training takes approximately 16 hours in total on a single RTX~4090. During inference, we use a CFG scale of 2.0.

\subsection{Evaluation Protocols and Metrics}

\noindent\textbf{Tri-Modal Motion Evaluator.}
As illustrated in Fig.~\ref{fig:evaluator_main}, to evaluate cross-modal alignment and kinematic realism, we construct a tri-modal evaluator following standard text-to-motion evaluation protocols~\cite{guo2022generating}. The evaluator encodes motion joint positions $\text{FK}(\mathbf{x})$, text $\mathbf{e}_t$, and spatial audio $\mathbf{e}_a$ (including the initial DOA) using modality-specific Transformer encoders and projects them into a shared embedding space. For multimodal inputs, a learned per-dimension gate fuses the text and acoustic embeddings into a unified condition representation,
\[
\mathbf{c} =
\text{MLP}\left(
\mathbf{g}\odot\mathbf{e}_a+
(\mathbf{1}-\mathbf{g})\odot\mathbf{e}_t
\right),
\mathbf{g}=
\sigma\left(
\text{MLP}\left([\mathbf{e}_t\|\mathbf{e}_a]\right)
\right).
\]
For text-only samples, it reduces to text-only condition. The evaluator is trained end-to-end using a symmetric InfoNCE objective~\cite{oord2018representation, radford2021learning}, where $s_{ij}=\mathbf{x}_i^\top\mathbf{c}_j$ denotes cosine similarity and $\tau=0.07$ is the temperature. Complete evaluator specifications are provided in Appendix~\ref{app:evaluator}.

\noindent\textbf{Metrics.}
Using the learned shared embedding space, we introduce two tri-modal retrieval metrics alongside the standard distribution-based measure FID:
\begin{itemize}
    \item \textbf{Condition-Motion Retrieval (CMR):} Given a multimodal (text, spatial-audio) query, measures whether the corresponding motion clip is retrieved from a candidate pool of $32$ samples.
    \item \textbf{Motion-Condition Retrieval (MCR):} Given a motion sequence, measures whether its corresponding multimodal condition is retrieved from $32$ candidate conditions.
\end{itemize}
We report R@$k$ (\textbf{CMR@$k$} and \textbf{MCR@$k$}) for $k\in\{1,2,3,5\}$ to measure cross-modal alignment, and \textbf{Fréchet Inception Distance (FID)}~\cite{heusel2017gans} to assess the overall distributional quality of the generated motion.

\begin{table}[htbp]
\centering
\setlength{\tabcolsep}{1pt}
\renewcommand{\arraystretch}{1}

\resizebox{0.49\textwidth}{!}{%
\begin{tabular}{@{}l|ccc|ccc|ccc@{}}
\toprule
\multirow{2}{*}{Method}
& \multicolumn{3}{c|}{Mixed}
& \multicolumn{3}{c|}{STAM}
& \multicolumn{3}{c}{HumanML3D} \\
\cmidrule(lr){2-4}
\cmidrule(lr){5-7}
\cmidrule(lr){8-10}
& FID$\downarrow$ & c2m R@1$\uparrow$ & m2c R@1$\uparrow$
& FID$\downarrow$ & c2m R@1$\uparrow$ & m2c R@1$\uparrow$
& FID$\downarrow$ & c2m R@1$\uparrow$ & m2c R@1$\uparrow$ \\
\midrule

GT
& 0.01 & 52.79 & 51.47
& 0.01 & 40.18 & 40.00
& 0.01 & 56.22 & 54.56 \\

\midrule

MoMask~\cite{guo2024momask}
& 1.09 & 44.40 & 42.81
& 2.06 & 25.36 & 25.09
& 1.45 & 49.40 & 47.76 \\

T2M-GPT~\cite{zhang2023generating}
& 0.40 & 43.32 & 41.67
& \underline{0.75} & 19.73 & 20.18
& 0.48 & 49.90 & 47.68 \\

TM2D~\cite{gong2023tm2d}
& 1.15 & 12.22 & 11.64
& 1.03 & 2.86 & 2.95
& 1.53 & 14.97 & 14.06 \\

EDGE~\cite{tseng2023edge}
& 3.90 & 29.93 & 28.46
& 1.03 & 34.73 & 33.66
& 5.69 & 28.58 & 26.79 \\

\midrule

\name
& \textbf{0.35} & \textbf{56.96} & \textbf{55.74}
& \textbf{0.73} & \underline{37.05} & \underline{37.95}
& \textbf{0.44} & \textbf{62.42} & \textbf{60.43} \\

\name\textsuperscript{$\dagger$}
& \underline{0.38} & \underline{54.38} & \underline{53.75}
& \underline{0.75} & \textbf{37.50} & \textbf{39.02}
& \underline{0.45} & \underline{58.97} & \underline{57.71} \\

\bottomrule
\end{tabular}%
}

\vspace{-2mm}
\caption{
\textbf{Quantitative comparison with baseline models on the mixed dataset of \dbname and HumanML3D.}
All methods are evaluated under identical protocols using our tri-modal evaluator.
\name and its minibatch-OT variant achieve state-of-the-art performance across the evaluated settings.
\textsuperscript{$\dagger$} denotes training with minibatch OT.
Best results are in \textbf{bold}; second-best are \underline{underlined}.
($\downarrow$: lower is better, $\uparrow$: higher is better).
}
\label{tab:comparison}
\vspace{-6mm}
\end{table}

\subsection{Main Comparison}

\noindent
\textbf{Qualitative Results.}
We compare \name~with T2M-GPT, MoMask, EDGE, and TM2D~\cite{zhang2023generating, guo2024momask, tseng2023edge, gong2023tm2d}. As shown in Fig.~\ref{fig:qualitative}, \name~produces more realistic and condition-responsive motions when jointly conditioned on text and spatial audio. We focus on FID and R-precision in the main paper; additional qualitative and quantitative results are provided in Appendix~\ref{app:experiments}.

\noindent
\textbf{Quantitative Results.}
We evaluate all methods using FID and our proposed CMR and MCR metrics, which extend R-precision to multimodal condition--motion retrieval. CMR measures whether a multimodal condition correctly retrieves its generated motion, while MCR evaluates the reverse retrieval from a motion sequence to its corresponding multimodal condition. We compare \name~against autoregressive~\cite{zhang2023generating} and non-autoregressive~\cite{guo2024momask, tseng2023edge, gong2023tm2d} baselines, covering both codebook-based~\cite{zhang2023generating, guo2024momask, gong2023tm2d} and diffusion-based~\cite{tseng2023edge} approaches. \name~achieves the best overall performance, with the highest R-precision scores across retrieval settings and the lowest FID. The superior R-precision demonstrates stronger alignment between generated motions and multimodal conditions, while the lower FID indicates closer agreement with the ground-truth motion distribution.

% ---------------------------------------------------------------------
% Combined ablation table: conditioning + capacity
% ---------------------------------------------------------------------
\begin{table*}[htbp]
\centering
\setlength{\tabcolsep}{3pt}
\renewcommand{\arraystretch}{1.2}

% ====================== Left: Conditioning Ablation ====================
\begin{subtable}[t]{0.495\textwidth}
\vspace{0pt}
\centering
\resizebox{\linewidth}{!}{%
\begin{tabular}{@{}l|ccc|ccc|ccc@{}}
\toprule
\multirow{2}{*}{Method}
& \multicolumn{3}{c|}{Mixed}
& \multicolumn{3}{c|}{STAM}
& \multicolumn{3}{c}{HumanML3D} \\
\cmidrule(lr){2-4}
\cmidrule(lr){5-7}
\cmidrule(lr){8-10}
& FID$\downarrow$ & c2m R@1$\uparrow$ & m2c R@1$\uparrow$
& FID$\downarrow$ & c2m R@1$\uparrow$ & m2c R@1$\uparrow$
& FID$\downarrow$ & c2m R@1$\uparrow$ & m2c R@1$\uparrow$ \\
\midrule
GT
& 0.01 & 52.79 & 51.47
& 0.01 & 40.18 & 40.00
& 0.01 & 56.22 & 54.56 \\
\midrule
\name
& \textbf{0.35} & 56.96 & 55.74
& \underline{0.73} & 37.05 & 37.95
& \underline{0.44} & \underline{62.42} & \underline{60.43} \\

~ GloVe words
& \textbf{0.35} & \textbf{58.39} & \textbf{56.58}
& 0.81 & \underline{38.57} & 38.48
& \textbf{0.43} & \textbf{63.91} & \textbf{61.49} \\

~ sentence-only
& 0.43 & 51.73 & 50.49
& 1.02 & 31.79 & 32.32
& 0.51 & 57.23 & 55.65 \\

~ raw audio
& 0.39 & \underline{57.06} & \underline{55.78}
& 0.91 & \textbf{40.80} & \underline{38.75}
& 0.46 & 61.34 & 60.23 \\

~ genre on
& \underline{0.37} & 56.35 & 54.68
& \textbf{0.69} & 34.46 & 35.18
& 0.46 & 62.35 & 60.01 \\

~ minibatch OT
& 0.38 & 54.38 & 53.75
& 0.75 & 37.50 & \textbf{39.02}
& 0.45 & 58.97 & 57.71 \\
\bottomrule
\end{tabular}%
}
\caption{\textbf{Conditioning components and minibatch OT.} Each variant modifies a single component under identical capacity and training iterations.}
\vspace{-2mm}
\label{tab:ablation_cond}
\end{subtable}
\hfill
% ========================= Right: Model Capacity =========================
\begin{subtable}[t]{0.495\textwidth}
\vspace{0pt}
\centering
\resizebox{\linewidth}{!}{%
\begin{tabular}{@{}lc|ccc|ccc|ccc@{}}
\toprule
\multirow{2}{*}{Depth} & \multirow{2}{*}{Width}
& \multicolumn{3}{c|}{Mixed}
& \multicolumn{3}{c|}{STAM}
& \multicolumn{3}{c}{HumanML3D} \\
\cmidrule(lr){3-5}
\cmidrule(lr){6-8}
\cmidrule(lr){9-11}
& & FID$\downarrow$ & c2m R@1$\uparrow$ & m2c R@1$\uparrow$
& FID$\downarrow$ & c2m R@1$\uparrow$ & m2c R@1$\uparrow$
& FID$\downarrow$ & c2m R@1$\uparrow$ & m2c R@1$\uparrow$ \\
\midrule
\multicolumn{2}{c}{GT}
& 0.01 & 52.79 & 51.47
& 0.01 & 40.18 & 40.00
& 0.01 & 56.22 & 54.56 \\
\midrule
8  & 512
& \underline{0.37} & 55.64 & 54.68
& 0.76 & 36.96 & \underline{36.70}
& \underline{0.45} & 60.76 & 59.58 \\

12 & 384
& 0.38 & 55.42 & 53.79
& \underline{0.71} & 35.62 & 35.45
& 0.48 & 60.99 & 58.77 \\

12 & 512
& \textbf{0.35} & \underline{56.96} & \textbf{55.74}
& 0.73 & \underline{37.05} & \textbf{37.95}
& \textbf{0.44} & \underline{62.42} & \textbf{60.43} \\

12 & 768
& 0.38 & 56.78 & 53.71
& \textbf{0.58} & 36.34 & 34.29
& 0.49 & 62.30 & 58.97 \\

16 & 512
& 0.39 & \textbf{57.51} & \underline{54.85}
& 0.73 & \textbf{38.21} & \underline{36.70}
& 0.48 & \textbf{62.83} & \underline{59.93} \\
\bottomrule
\end{tabular}%
}
\caption{\textbf{Model capacity.} The proposed $12\times512$ configuration provides a good balance between FID and R-precision.}
\vspace{-2mm}
\label{tab:ablation_size}
\end{subtable}

\vspace{-1mm}
\caption{\textbf{Ablation studies on model design.}
\textbf{a)} Ablation of conditioning components and minibatch-OT training.
The proposed configuration \name achieves a balanced trade-off between kinematic quality (FID) and semantic alignment (R-precision).
\textbf{(b)} Model-capacity ablation, where $12\times512$ is the proposed configuration used in subsequent comparisons.
Best results are in \textbf{bold}; second-best are \underline{underlined}.}
\label{tab:ablation}
\vspace{-4mm}
\end{table*}

\subsection{Ablation Study}

We conduct ablation studies to analyze the contribution of different conditioning components and training strategies. All variants are trained with identical hyper-parameters and differ from the full model by only one modified component.

\noindent\textbf{Conditioning Mechanisms.}
Table~\ref{tab:ablation_cond} summarizes the effects of different text, audio, and spatial conditioning designs. Our full configuration achieves the best overall trade-off between motion quality (FID) and cross-modal alignment (R-precision).

Replacing the CLIP word-level contextual embeddings with GloVe embeddings (Row 3 vs.\ Row 2) improves several retrieval metrics but substantially degrades FID, indicating degraded motion distribution quality. Removing fine-grained word-level features and using only the global CLIP sentence embedding (Row 4 vs.\ Row 2) consistently reduces performance, demonstrating the importance of token-level semantic information for capturing detailed motion variations.

Replacing CLAP embeddings with low-level acoustic features of the same temporal resolution also degrades FID across all evaluation subsets, although it achieves slightly higher R-precision on the mixed and \dbname test sets. This suggests that high-level acoustic representations provide more effective motion-quality guidance, while low-level features may preserve certain retrieval-related cues.

We further evaluate the genre token introduced in~\cite{xu2026mospa}. Activating the genre token provides no consistent improvement, with only a marginal FID gain on \dbname accompanied by slightly lower R-precision. We hypothesize that genre information is partially redundant with textual descriptions, which already contain attributes such as adjectives and adverbs describing motion style and intensity.

\noindent\textbf{Training Strategy.}
We additionally evaluate the effect of minibatch Optimal Transport (OT)~\cite{tong2023improving, pooladian2023multisample}. As shown in Table~\ref{tab:ablation_cond}, applying minibatch OT provides only marginal improvements on R-precision for \dbname and does not consistently improve overall performance, suggesting limited benefits under our flow matching formulation.

\noindent\textbf{Model Capacity and Scaling.}
Row 4 corresponds to the proposed configuration used throughout our comparisons and ablation studies. We further investigate the effect of network capacity on motion generation quality and semantic alignment. As shown in Table~\ref{tab:ablation_size}, reducing the model capacity, particularly the number of Transformer layers, leads to a slight performance degradation (Rows 2--3 vs.\ Row 4). In contrast, increasing the model size beyond the proposed configuration provides only marginal improvements across evaluation metrics (Rows 5--6 vs.\ Row 4), suggesting that the proposed configuration achieves a favorable accuracy-efficiency trade-off. Additional experiments and details are provided in Appendix~\ref{app:experiments}.

\subsection{\name for Simulated Humanoid Control}

\begin{figure}[htbp]
    \centering
    \includegraphics[width=0.9\linewidth]{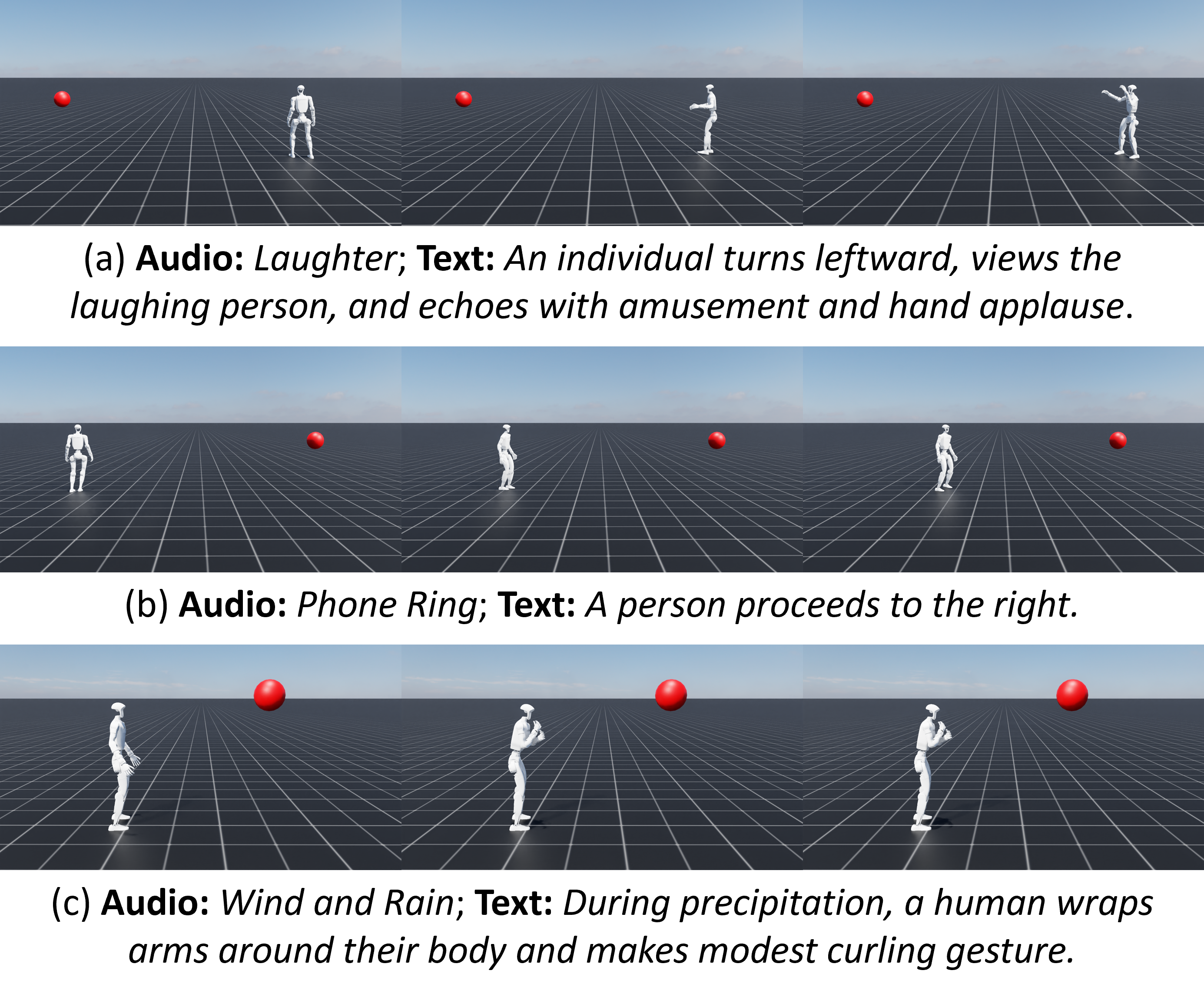}
    \vspace{-4mm}
    \caption{\textbf{\name for simulated humanoid control.} 
    The Unitree G1 humanoid in simulation tracks three motions generated by \name from paired spatial audio and text inputs, with the red sphere indicating the audio source direction. These examples demonstrate that our multimodal motion generator can serve as an effective high-level planner for sound- and language-guided humanoid behavior.
    }
    \label{fig:robot}
    \vspace{-6mm}
\end{figure}

To validate the applicability of \name to humanoid motion control, we deploy it to control a Unitree G1 humanoid robot using the Orbit environment~\cite{mittal2023orbit}. Specifically, we perform motion tracking on the G1 robot using the state-of-the-art BeyondMimic~\cite{liao2026beyondmimic} framework. The generated 3D human motions are used as reference motions for the robot to track. We first perform inverse kinematics to obtain the robot poses using the GMR~\cite{araujo2025retargeting} retargeting framework. We show that the robot is tracking three different sound- and text-conditioned motions in Fig.~\ref{fig:robot}, which illustrates that \name can serve as an effective high-level planner for humanoid robots.

\section{Conclusion}
\label{sec:conclusion}
We study a new task for human motion synthesis jointly conditioned on spatial audio and natural language, and curate \textbf{\dbname}, a benchmark of motion sequences paired with spatial audio and motion-aligned textual captions for multi-conditioned motion generation and evaluation. We further propose \textbf{\name}, the first framework tailored for fusing spatial audio and text for motion generation. \name uses a Motion VAE for compact motion latents and a transformer-based flow matching model for motion synthesis. We also develop a tri-modal retrieval evaluator, with contrastively score alignment between motion, audio, and text for evaluation. Experiments across datasets show that \name achieves SOTA performance, producing temporally coherent, spatially grounded, and semantically faithful motions.\\
\noindent\textbf{Limitations and future work.} Although \name achieves higher generation quality compared with baselines, its performance can degrade under strong audio noise or temporal misalignment between modalities.
Extending our framework to multi-person interaction~\cite{wu2026text2interact, javed2024intermask,liang2024intergen} and scene- and affordance-aware control~\cite{cong2024laserhuman, wang2024move, yi2024generating} is an important direction. Another avenue is to incorporate stronger physical priors, such as simulated humanoids~\cite{peng2022ase, peng2018deepmimic} and more explicit contact modeling, to further improve realism. Finally, developing efficient and robust streaming generation~\cite{xiao2025motionstreamer} would help deploy spatial-audio-and-text-conditioned motion synthesis in real-time interactive settings.
\clearpage
{
    \small
    \bibliographystyle{ieeenat_fullname}
    \bibliography{main}
}

\clearpage
\appendix
\renewcommand\thefigure{\Alph{section}\arabic{figure}}    
\renewcommand\thetable{\Alph{section}\arabic{table}}

\section{Motion Features}
\setcounter{table}{0}
\label{app:motion_features}

We present the complete list of motion features and their corresponding dimensions in Table~\ref{tab:motion_features}. Our motion vector structure is largely inspired by~\cite{xiao2025motionstreamer}. Consequently, the shape of the motion vectors $\mathbf{x}$ becomes $\mathbf{x} \in \mathbb{R}^{T_f \times 272}$, where $T_f$ denotes the length of the motion.

\begin{table}[!ht]
  \centering
  \setlength{\tabcolsep}{8pt}
  \renewcommand{\arraystretch}{1.2}
  \begin{tabular}{l | c}
      \hline
      \textbf{Feature Name} & \textbf{Dimension} \\
      \hline
      Root Linear Velocity on XZ plane & 2 \\
      Root Heading Rotation & 6 \\
      Joint Positions (22 $\times$ 3) & 66 \\
      Joint Velocities (22 $\times$ 3) & 66 \\
      Joint Rotations (22 $\times$ 6) & 132 \\
      \midrule
      Total & 272 \\
      \hline
  \end{tabular}
  \vspace{2mm}
  \caption{Motion features and their dimensions. Each frame of motion is represented by a 272-dimensional vector.}
  \label{tab:motion_features}
\end{table}

\section{Multimodal Evaluator}
\label{app:evaluator}

\subsection{Audio Features}
For a fair evaluation of audio--motion alignment, we represent the semantic content of each audio clip using low-level acoustic features extracted directly from the raw waveform, rather than pretrained CLAP embeddings~\cite{elizalde2023clap}. Specifically, we concatenate a $256$D Mel-spectrogram representation with its corresponding $256$D temporal delta features, resulting in a $512$D acoustic representation. The features are extracted at $7.5$~FPS to match the temporally strided motion representation used by the evaluator. The detailed feature composition is summarized in Table~\ref{tab:audio_features_eval}.

\begin{table}[!ht]
    \centering
    \setlength{\tabcolsep}{8pt}
    \renewcommand{\arraystretch}{1.2}
    \begin{tabular}{l | c}
        \hline
        \textbf{Feature} & \textbf{Dimension} \\
        \hline
        Mel-spectrogram & 256 \\
        Temporal delta features & 256 \\
        \midrule
        Total & 512 \\
        \hline
    \end{tabular}
    \vspace{2mm}
    \caption{Composition of the acoustic features used by the evaluator. The concatenated acoustic representation has a total dimension of $512$.}
    \label{tab:audio_features_eval}
\end{table}

\subsection{Evaluator Architecture}

To evaluate cross-modal alignment and motion quality, we construct a multimodal co-embedding evaluator that projects motion, text, and spatial-audio inputs into a shared $128$D embedding space. After training, the evaluator is frozen and used to compute the retrieval-based metrics and FID reported in our experiments.

\noindent\textbf{Input Streams.}
All temporal modalities are represented at $7.5$~FPS:
\begin{itemize}
    \item \textbf{Motion:} The original $272$D local motion representation is temporally strided to $7.5$~FPS and converted through forward kinematics into $66$D global joint positions corresponding to $22$ joints.

    \item \textbf{Text:} Each caption is represented by up to $30$ word tokens using $300$D GloVe embeddings, augmented with projected $16$D positional one-hot encodings.

    \item \textbf{Audio and DOA:} Each audio clip is represented by the $512$D acoustic features described above and normalized using z-score normalization. The initial $3$D DOA vector is projected through an MLP bottleneck and incorporated into the acoustic representation.
\end{itemize}

\noindent\textbf{Architecture and Dynamic Fusion.}
Each modality is processed by a separate $3$-layer Transformer encoder with width $256$, $4$ attention heads, GELU activations, and dropout of $0.1$. A prepended $\mathtt{[CLS]}$ token summarizes each sequence, and its output is mapped to the shared embedding space through a projection head consisting of a linear layer, Layer Normalization, LeakyReLU activation, and a final projection to $128$ dimensions.

The text and spatial-audio embeddings are subsequently fused using a learned sigmoid gate to obtain a unified condition representation. The gate adaptively balances the contributions of the text and audio embeddings on a per-dimension basis. For samples without audio, the condition representation is directly set to the text embedding.

\noindent\textbf{Training Protocol.}
The evaluator is trained on the combined training splits of HumanML3D and \dbname using a symmetric contrastive objective between motion and condition embeddings. Given a batch of $N$ matched motion--condition pairs, we optimize the symmetric InfoNCE loss~\cite{oord2018representation, radford2021learning}:
\begin{multline}
    \mathcal{L}_{\text{NCE}} =
    -\frac{1}{N}\sum_{i=1}^{N}
    \biggl[
    \log
    \frac{\exp(s_{ii}/\tau)}
    {\sum_{j\in\mathcal{P}_i}\exp(s_{ij}/\tau)}
    + \\
    \log
    \frac{\exp(s_{ii}/\tau)}
    {\sum_{j:(j,i)\in\mathcal{P}}\exp(s_{ji}/\tau)}
    \biggr],
\end{multline}
where $s_{ij}=\mathbf{x}_i^\top\mathbf{c}_j$ denotes the cosine similarity between the normalized motion embedding $\mathbf{x}_i$ and condition embedding $\mathbf{c}_j$, $P_i=\{j:(i,j)\text{ is not masked}\}$ and $\tau=0.07$ is the temperature parameter.

\begin{itemize}
    \item \textbf{False-Negative Masking:} For each pair of samples, we compute the cosine similarity between their mean-pooled input caption embeddings, encoded audio embeddings, and DOA vectors. The caption similarity is computed from the length-masked mean of the input GloVe features before the text Transformer. A pair is excluded from the contrastive denominators only when all applicable similarities exceed $0.95$ (i.e., text, audio, and DOA for samples with audio; text only for HumanML3D samples without audio).

    \item \textbf{Data Augmentation:} During training, we apply random temporal cropping to motion sequences, contiguous temporal masking over $15\%$ of the audio sequence, and Gaussian perturbations with $\sigma=0.05$ to the DOA vectors.

    \item \textbf{Optimization Details:} The evaluator is optimized using AdamW with an initial learning rate of $2\times10^{-4}$, weight decay of $0.05$, and a batch size of $256$. Training is performed for $300$ epochs with step-wise learning-rate decay at epochs $150$ and $250$. We select the checkpoint with the highest mean $R@1$ on the validation set, evaluated every $10$ epochs.
\end{itemize}

\section{Extended Experiments and Analysis}
\setcounter{table}{0}
\label{app:experiments}

In this section, we provide extended baseline comparisons and ablation analyses using additional evaluation metrics, including Fr\'echet Inception Distance (FID)~\cite{heusel2017gans}, Top-$k$ R-Precision, Mean Median Retrieval Rank (MedR), Diversity, and Average Pairwise Distance (APD)~\cite{dou2023c}. All models are trained on the joint mixed dataset, while evaluation is performed separately on three test splits:
\begin{enumerate}
    \item \textbf{Joint Co-Training Test Set:} The combined test split of \dbname and HumanML3D~\cite{guo2022generating}, containing 5,088 samples, which evaluates overall performance under the joint training setting.
    
    \item \textbf{\dbname Test Set:} The \dbname test split containing 1,120 samples, which focuses on spatial-audio- and text-conditioned motion synthesis.
    
    \item \textbf{HumanML3D Test Set:} The standalone HumanML3D test split~\cite{guo2022generating}, containing 3,968 samples, which evaluates text-conditioned motion generation.
\end{enumerate}
For completeness, we report results on all three evaluation splits for each set of experiments.

We compute Diversity and Average Pairwise Distance (APD) on the corresponding motion feature representations $\mathbf{x}$. Diversity measures the average distance between randomly paired generated samples:
\begin{equation}
    \mathrm{Diversity}
    =
    \frac{1}{K}
    \sum_{k=1}^{K}
    \left\|
    \mathbf{x}_{a_k} - \mathbf{x}_{b_k}
    \right\|_2,
    \qquad
    K=\min(300,N),
    \label{eq:div}
\end{equation}
where $(a_k,b_k)$ denotes a randomly selected pair of distinct samples.

APD measures the average pairwise distance across all generated samples:
\begin{equation}
    \mathrm{APD}
    =
    \frac{2}{N(N-1)}
    \sum_{i=1}^{N}
    \sum_{j=i+1}^{N}
    \left\|
    \mathbf{x}_i-\mathbf{x}_j
    \right\|_2.
    \label{eq:apd}
\end{equation}

\begin{table*}[!ht]
\centering
\small
\setlength{\tabcolsep}{5pt}
\begin{tabular}{@{}l c c c c c c c c c c c cc@{}}
\toprule
\multirow{2}{*}{\textbf{Method}} & \multirow{2}{*}{\textbf{FID}$\downarrow$}
    & \multicolumn{5}{c}{\textbf{C2M}}
    & \multicolumn{5}{c}{\textbf{M2C}}
    & \multirow{2}{*}{\textbf{Div}$\rightarrow$} & \multirow{2}{*}{\textbf{APD}$\rightarrow$} \\
\cmidrule(lr){3-7}\cmidrule(lr){8-12}
    & & \textbf{R@1}$\uparrow$ & \textbf{R@2}$\uparrow$ & \textbf{R@3}$\uparrow$ & \textbf{R@5}$\uparrow$ & \textbf{MedR}$\downarrow$
      & \textbf{R@1}$\uparrow$ & \textbf{R@2}$\uparrow$ & \textbf{R@3}$\uparrow$ & \textbf{R@5}$\uparrow$ & \textbf{MedR}$\downarrow$ & \\
\midrule
GT & 0.01 & 52.79 & 68.69 & 77.14 & 85.89 & 1.33 & 51.47 & 68.69 & 77.14 & 85.36 & 1.38 & 35.84 & 37.51 \\
\midrule
MoMask & 1.09 & 44.40 & 60.99 & 68.97 & 78.71 & 1.85 & 42.81 & 59.93 & 68.71 & 79.19 & 1.86 & \textbf{35.52} & \textbf{38.05} \\
T2M-GPT & 0.40 & 43.32 & 57.63 & 65.62 & 76.12 & 1.96 & 41.67 & 56.45 & 65.35 & 75.49 & 2.06 & \underline{34.40} & 35.62 \\
TM2D & 1.15 & 12.22 & 19.58 & 26.16 & 35.40 & 9.45 & 11.64 & 19.38 & 25.28 & 34.73 & 9.77 & 33.06 & 32.80 \\
EDGE & 3.90 & 29.93 & 44.99 & 54.17 & 66.14 & 3.25 & 28.46 & 43.97 & 53.73 & 64.78 & 3.48 & 31.30 & 33.01 \\
\midrule
\textbf{\name} & \textbf{0.35} & \textbf{56.96} & \textbf{72.62} & \textbf{80.78} & \textbf{88.21} & \underline{1.30} & \textbf{55.74} & \textbf{72.23} & \textbf{79.91} & \textbf{88.05} & \underline{1.30} & 33.93 & \underline{36.23} \\
\textbf{\name + OT} & \underline{0.38} & \underline{54.38} & \underline{70.30} & \underline{78.97} & \underline{87.24} & \textbf{1.28} & \underline{53.75} & \underline{70.40} & \underline{78.73} & \underline{87.07} & \textbf{1.28} & 33.17 & 34.98 \\
\bottomrule
\end{tabular}
\caption{Mixed test set (HumanML3D + \dbname, $N{=}5088$). $\downarrow$ = lower is better, $\uparrow$ = higher is better; $\rightarrow$ (Div/APD): closer to GT is better. Best in \textbf{bold}, second best \underline{underlined}.}
\label{tab:compare_mixed}
\end{table*}

\begin{table*}[!ht]
\centering
\small
\setlength{\tabcolsep}{5pt}
\begin{tabular}{@{}l c c c c c c c c c c c cc@{}}
\toprule
\multirow{2}{*}{\textbf{Method}} & \multirow{2}{*}{\textbf{FID}$\downarrow$}
    & \multicolumn{5}{c}{\textbf{C2M}}
    & \multicolumn{5}{c}{\textbf{M2C}}
    & \multirow{2}{*}{\textbf{Div}$\rightarrow$} & \multirow{2}{*}{\textbf{APD}$\rightarrow$} \\
\cmidrule(lr){3-7}\cmidrule(lr){8-12}
    & & \textbf{R@1}$\uparrow$ & \textbf{R@2}$\uparrow$ & \textbf{R@3}$\uparrow$ & \textbf{R@5}$\uparrow$ & \textbf{MedR}$\downarrow$
      & \textbf{R@1}$\uparrow$ & \textbf{R@2}$\uparrow$ & \textbf{R@3}$\uparrow$ & \textbf{R@5}$\uparrow$ & \textbf{MedR}$\downarrow$ & \\
\midrule
GT & 0.01 & 40.18 & 59.64 & 70.54 & 84.64 & 2.00 & 40.00 & 63.84 & 76.16 & 87.32 & 1.80 & 39.69 & 40.41 \\
\midrule
MoMask & 2.06 & 25.36 & 43.04 & 53.48 & 69.38 & 3.14 & 25.09 & 42.77 & 54.20 & 72.59 & 3.14 & \underline{40.18} & \textbf{40.55} \\
T2M-GPT & \underline{0.75} & 19.73 & 35.18 & 46.79 & 65.36 & 3.69 & 20.18 & 35.00 & 47.32 & 65.89 & 3.60 & 37.78 & 38.36 \\
TM2D & 1.03 & 2.86 & 5.71 & 9.46 & 15.71 & 15.89 & 2.95 & 5.71 & 8.75 & 15.45 & 15.80 & 36.06 & 36.70 \\
EDGE & 1.03 & 34.73 & 54.29 & 67.68 & 82.23 & 2.17 & 33.66 & 55.71 & 68.48 & 83.93 & \underline{2.14} & \textbf{39.68} & \underline{39.94} \\
\midrule
\textbf{\name} & \textbf{0.73} & \underline{37.05} & \textbf{58.75} & \textbf{71.96} & \textbf{85.98} & \underline{2.11} & \underline{37.95} & \textbf{61.25} & \textbf{73.84} & \textbf{87.95} & \textbf{2.03} & 36.34 & 37.48 \\
\textbf{\name + OT} & \underline{0.75} & \textbf{37.50} & \underline{57.41} & \underline{70.80} & \underline{85.62} & \textbf{2.09} & \textbf{39.02} & \underline{59.38} & \underline{72.59} & \underline{87.14} & \textbf{2.03} & 35.83 & 37.04 \\
\bottomrule
\end{tabular}
\caption{\dbname test set ($N{=}1120$). $\downarrow$ = lower is better, $\uparrow$ = higher is better; $\rightarrow$ (Div/APD): closer to GT is better. Best in \textbf{bold}, second best \underline{underlined}.}
\label{tab:compare_stam}
\end{table*}

\begin{table*}[!ht]
\centering
\small
\setlength{\tabcolsep}{5pt}
\begin{tabular}{@{}l c c c c c c c c c c c cc@{}}
\toprule
\multirow{2}{*}{\textbf{Method}} & \multirow{2}{*}{\textbf{FID}$\downarrow$}
    & \multicolumn{5}{c}{\textbf{C2M}}
    & \multicolumn{5}{c}{\textbf{M2C}}
    & \multirow{2}{*}{\textbf{Div}$\rightarrow$} & \multirow{2}{*}{\textbf{APD}$\rightarrow$} \\
\cmidrule(lr){3-7}\cmidrule(lr){8-12}
    & & \textbf{R@1}$\uparrow$ & \textbf{R@2}$\uparrow$ & \textbf{R@3}$\uparrow$ & \textbf{R@5}$\uparrow$ & \textbf{MedR}$\downarrow$
      & \textbf{R@1}$\uparrow$ & \textbf{R@2}$\uparrow$ & \textbf{R@3}$\uparrow$ & \textbf{R@5}$\uparrow$ & \textbf{MedR}$\downarrow$ & \\
\midrule
GT & 0.01 & 56.22 & 71.17 & 78.78 & 85.89 & 1.17 & 54.56 & 69.91 & 77.04 & 84.75 & 1.25 & 34.92 & 35.92 \\
\midrule
MoMask & 1.45 & 49.40 & 65.70 & 73.08 & 80.97 & 1.51 & 47.76 & 64.29 & 72.25 & 80.80 & \underline{1.56} & \underline{35.80} & \textbf{36.66} \\
T2M-GPT & 0.48 & 49.90 & 63.81 & 70.74 & 78.98 & 1.44 & 47.68 & 62.35 & 70.21 & 78.00 & 1.63 & 33.66 & 33.97 \\
TM2D & 1.53 & 14.97 & 23.39 & 31.07 & 40.88 & 7.56 & 14.06 & 23.21 & 29.86 & 39.97 & 7.90 & 30.93 & 30.70 \\
EDGE & 5.69 & 28.58 & 42.24 & 50.10 & 61.16 & 3.55 & 26.79 & 40.27 & 49.12 & 59.00 & 3.84 & 30.03 & 30.17 \\
\midrule
\textbf{\name} & \textbf{0.44} & \textbf{62.42} & \textbf{76.29} & \textbf{82.84} & \textbf{88.53} & \underline{1.10} & \textbf{60.43} & \textbf{75.05} & \textbf{81.17} & \textbf{87.80} & \textbf{1.10} & \textbf{34.60} & \underline{35.00} \\
\textbf{\name + OT} & \underline{0.45} & \underline{58.97} & \underline{73.79} & \underline{80.92} & \underline{87.55} & \textbf{1.07} & \underline{57.71} & \underline{72.98} & \underline{79.84} & \underline{86.82} & \textbf{1.10} & 32.92 & 33.54 \\
\bottomrule
\end{tabular}
\caption{HumanML3D test set ($N{=}3968$). $\downarrow$ = lower is better, $\uparrow$ = higher is better; $\rightarrow$ (Div/APD): closer to GT is better. Best in \textbf{bold}, second best \underline{underlined}.}
\label{tab:compare_hml}
\end{table*}

\subsection{Quantitative Comparison}

As shown in Tab.~\ref{tab:compare_mixed}, \ref{tab:compare_stam}, and \ref{tab:compare_hml}, \name~or \name+Optimal Transport (OT) consistently achieves the strongest overall performance across the three evaluation splits and nearly all reported metrics.

The main exceptions are Diversity and APD, for which some baseline methods obtain higher numerical scores. However, higher Diversity and APD do not necessarily correspond to better motion quality or condition alignment. In particular, methods such as MoMask~\cite{guo2024momask} exhibit higher diversity at the expense of FID and R-Precision, whereas \name~provides a more favorable balance between motion quality, diversity, and multimodal alignment. The results suggest that \name~generates diverse motions without sacrificing motion realism or multimodal condition alignment.

\begin{table*}[!ht]
\centering
\small
\setlength{\tabcolsep}{5pt}
\begin{tabular}{@{}l c c c c c c c c c c c cc@{}}
\toprule
\multirow{2}{*}{\textbf{Method}} & \multirow{2}{*}{\textbf{FID}$\downarrow$}
    & \multicolumn{5}{c}{\textbf{C2M}}
    & \multicolumn{5}{c}{\textbf{M2C}}
    & \multirow{2}{*}{\textbf{Div}$\rightarrow$} & \multirow{2}{*}{\textbf{APD}$\rightarrow$} \\
\cmidrule(lr){3-7}\cmidrule(lr){8-12}
    & & \textbf{R@1}$\uparrow$ & \textbf{R@2}$\uparrow$ & \textbf{R@3}$\uparrow$ & \textbf{R@5}$\uparrow$ & \textbf{MedR}$\downarrow$
      & \textbf{R@1}$\uparrow$ & \textbf{R@2}$\uparrow$ & \textbf{R@3}$\uparrow$ & \textbf{R@5}$\uparrow$ & \textbf{MedR}$\downarrow$ & \\
\midrule
GT & 0.01 & 52.79 & 68.69 & 77.14 & 85.89 & 1.33 & 51.47 & 68.69 & 77.14 & 85.36 & 1.38 & 35.84 & 37.51 \\
\midrule
\name & \textbf{0.35} & 56.96 & \underline{72.62} & \underline{80.78} & \underline{88.21} & 1.30 & 55.74 & \underline{72.23} & 79.91 & 88.05 & 1.30 & 33.93 & \textbf{36.23} \\
~ GloVe words & \textbf{0.35} & \textbf{58.39} & \textbf{74.00} & \textbf{82.08} & \textbf{89.76} & \textbf{1.25} & \textbf{56.58} & \textbf{73.53} & \textbf{81.72} & \textbf{89.29} & \underline{1.27} & \underline{34.19} & \underline{36.13} \\
~ sentence-only & 0.43 & 51.73 & 68.12 & 76.55 & 84.93 & 1.42 & 50.49 & 67.22 & 75.88 & 85.38 & 1.45 & \underline{34.19} & \textbf{36.23} \\
~ raw audio & 0.39 & \underline{57.06} & 72.17 & 79.83 & 87.58 & \underline{1.26} & \underline{55.78} & 72.01 & \underline{79.93} & \underline{88.17} & \textbf{1.26} & \textbf{34.29} & 36.06 \\
~ genre on & \underline{0.37} & 56.35 & 71.27 & 79.23 & 87.81 & 1.31 & 54.68 & 71.29 & 79.40 & 87.78 & 1.30 & 33.80 & 36.08 \\
~ minibatch OT & 0.38 & 54.38 & 70.30 & 78.97 & 87.24 & 1.28 & 53.75 & 70.40 & 78.73 & 87.07 & 1.28 & 33.17 & 34.98 \\
\bottomrule
\end{tabular}
\caption{Condition ablation on the mixed test set (HumanML3D + \dbname, $N{=}5088$). $\downarrow$ = lower is better, $\uparrow$ = higher is better; $\rightarrow$ (Div/APD): closer to GT is better. Best in \textbf{bold}, second best \underline{underlined}.}
\label{tab:abl_cond_mixed}
\end{table*}

\begin{table*}[!ht]
\centering
\small
\setlength{\tabcolsep}{5pt}
\begin{tabular}{@{}l c c c c c c c c c c c cc@{}}
\toprule
\multirow{2}{*}{\textbf{Method}} & \multirow{2}{*}{\textbf{FID}$\downarrow$}
    & \multicolumn{5}{c}{\textbf{C2M}}
    & \multicolumn{5}{c}{\textbf{M2C}}
    & \multirow{2}{*}{\textbf{Div}$\rightarrow$} & \multirow{2}{*}{\textbf{APD}$\rightarrow$} \\
\cmidrule(lr){3-7}\cmidrule(lr){8-12}
    & & \textbf{R@1}$\uparrow$ & \textbf{R@2}$\uparrow$ & \textbf{R@3}$\uparrow$ & \textbf{R@5}$\uparrow$ & \textbf{MedR}$\downarrow$
      & \textbf{R@1}$\uparrow$ & \textbf{R@2}$\uparrow$ & \textbf{R@3}$\uparrow$ & \textbf{R@5}$\uparrow$ & \textbf{MedR}$\downarrow$ & \\
\midrule
GT & 0.01 & 40.18 & 59.64 & 70.54 & 84.64 & 2.00 & 40.00 & 63.84 & 76.16 & 87.32 & 1.80 & 39.69 & 40.41 \\
\midrule
\name & \underline{0.73} & 37.05 & \underline{58.75} & \textbf{71.96} & \textbf{85.98} & 2.11 & 37.95 & \textbf{61.25} & \textbf{73.84} & \textbf{87.95} & \underline{2.03} & \underline{36.34} & \textbf{37.48} \\
~ GloVe words & 0.81 & \underline{38.57} & \textbf{58.84} & \underline{71.79} & 85.45 & \underline{2.03} & 38.48 & 59.38 & 72.23 & 85.89 & \underline{2.03} & 36.21 & 36.87 \\
~ sentence-only & 1.02 & 31.79 & 52.50 & 65.54 & 79.64 & 2.34 & 32.32 & 53.48 & 67.41 & 84.29 & 2.20 & 34.07 & 35.25 \\
~ raw audio & 0.91 & \textbf{40.80} & 58.04 & 69.91 & 83.04 & \textbf{1.97} & \underline{38.75} & \underline{60.27} & 71.79 & 86.34 & \textbf{1.97} & 35.74 & 36.65 \\
~ genre on & \textbf{0.69} & 34.46 & 55.45 & 69.55 & 84.29 & 2.20 & 35.18 & 56.70 & 70.45 & 85.54 & 2.17 & \textbf{36.67} & \underline{37.41} \\
~ minibatch OT & 0.75 & 37.50 & 57.41 & 70.80 & \underline{85.62} & 2.09 & \textbf{39.02} & 59.38 & \underline{72.59} & \underline{87.14} & \underline{2.03} & 35.83 & 37.04 \\
\bottomrule
\end{tabular}
\caption{Condition ablation on the \dbname test set ($N{=}1120$). $\downarrow$ = lower is better, $\uparrow$ = higher is better; $\rightarrow$ (Div/APD): closer to GT is better. Best in \textbf{bold}, second best \underline{underlined}.}
\label{tab:abl_cond_stam}
\end{table*}

\begin{table*}[!ht]
\centering
\small
\setlength{\tabcolsep}{5pt}
\begin{tabular}{@{}l c c c c c c c c c c c cc@{}}
\toprule
\multirow{2}{*}{\textbf{Method}} & \multirow{2}{*}{\textbf{FID}$\downarrow$}
    & \multicolumn{5}{c}{\textbf{C2M}}
    & \multicolumn{5}{c}{\textbf{M2C}}
    & \multirow{2}{*}{\textbf{Div}$\rightarrow$} & \multirow{2}{*}{\textbf{APD}$\rightarrow$} \\
\cmidrule(lr){3-7}\cmidrule(lr){8-12}
    & & \textbf{R@1}$\uparrow$ & \textbf{R@2}$\uparrow$ & \textbf{R@3}$\uparrow$ & \textbf{R@5}$\uparrow$ & \textbf{MedR}$\downarrow$
      & \textbf{R@1}$\uparrow$ & \textbf{R@2}$\uparrow$ & \textbf{R@3}$\uparrow$ & \textbf{R@5}$\uparrow$ & \textbf{MedR}$\downarrow$ & \\
\midrule
GT & 0.01 & 56.22 & 71.17 & 78.78 & 85.89 & 1.17 & 54.56 & 69.91 & 77.04 & 84.75 & 1.25 & 34.92 & 35.92 \\
\midrule
\name & \underline{0.44} & \underline{62.42} & \underline{76.29} & \underline{82.84} & \underline{88.53} & 1.10 & \underline{60.43} & \underline{75.05} & 81.17 & 87.80 & 1.10 & \underline{34.60} & 35.00 \\
~ GloVe words & \textbf{0.43} & \textbf{63.91} & \textbf{78.30} & \textbf{84.75} & \textbf{90.62} & \textbf{1.03} & \textbf{61.49} & \textbf{77.29} & \textbf{83.92} & \textbf{90.05} & \underline{1.08} & 34.44 & \underline{35.07} \\
~ sentence-only & 0.51 & 57.23 & 72.63 & 79.71 & 86.06 & 1.18 & 55.65 & 70.79 & 78.02 & 85.51 & 1.23 & 35.34 & \textbf{35.66} \\
~ raw audio & 0.46 & 61.34 & 75.93 & 82.26 & \underline{88.53} & \underline{1.06} & 60.23 & 74.92 & \underline{81.60} & \underline{88.43} & 1.09 & \textbf{34.64} & 35.04 \\
~ genre on & 0.46 & 62.35 & 75.43 & 81.80 & 88.36 & 1.09 & 60.01 & 74.72 & 81.53 & 88.13 & \textbf{1.07} & 34.41 & 34.85 \\
~ minibatch OT & 0.45 & 58.97 & 73.79 & 80.92 & 87.55 & 1.07 & 57.71 & 72.98 & 79.84 & 86.82 & 1.10 & 32.92 & 33.54 \\
\bottomrule
\end{tabular}
\caption{Condition ablation on the HumanML3D test set ($N{=}3968$). $\downarrow$ = lower is better, $\uparrow$ = higher is better; $\rightarrow$ (Div/APD): closer to GT is better. Best in \textbf{bold}, second best \underline{underlined}.}
\label{tab:abl_cond_hml}
\end{table*}

\subsection{Ablation on Conditioning}

As shown in Tab.~\ref{tab:abl_cond_mixed}, \ref{tab:abl_cond_stam}, and \ref{tab:abl_cond_hml}, the conditioning ablations reveal consistent trade-offs between motion quality and semantic alignment. While replacing the CLIP contextual word embeddings with GloVe features improves several metrics on the mixed and HumanML3D test sets, it substantially degrades FID on the \dbname test set. This indicates that the choice of textual representation affects both semantic alignment and motion distribution quality.

Overall, we select the proposed configuration as the default setting because it provides the most balanced performance across FID and R-Precision and avoids pronounced degradation on any individual evaluation split.

\begin{table*}[!ht]
\centering
\small
\setlength{\tabcolsep}{5pt}
\begin{tabular}{@{}l l c c c c c c c c c c c cc@{}}
\toprule
\multirow{2}{*}{\textbf{Depth}} & \multirow{2}{*}{\textbf{Width}} & \multirow{2}{*}{\textbf{FID}$\downarrow$}
    & \multicolumn{5}{c}{\textbf{C2M}}
    & \multicolumn{5}{c}{\textbf{M2C}}
    & \multirow{2}{*}{\textbf{Div}$\rightarrow$} & \multirow{2}{*}{\textbf{APD}$\rightarrow$} \\
\cmidrule(lr){4-8}\cmidrule(lr){9-13}
    & & & \textbf{R@1}$\uparrow$ & \textbf{R@2}$\uparrow$ & \textbf{R@3}$\uparrow$ & \textbf{R@5}$\uparrow$ & \textbf{MedR}$\downarrow$
      & \textbf{R@1}$\uparrow$ & \textbf{R@2}$\uparrow$ & \textbf{R@3}$\uparrow$ & \textbf{R@5}$\uparrow$ & \textbf{MedR}$\downarrow$ & \\
\midrule
GT & -- & 0.01 & 52.79 & 68.69 & 77.14 & 85.89 & 1.33 & 51.47 & 68.69 & 77.14 & 85.36 & 1.38 & 35.84 & 37.51 \\
\midrule
8 & 512 & \underline{0.37} & 55.64 & 70.34 & 78.71 & 87.38 & 1.32 & 54.68 & 71.05 & 79.62 & 87.24 & \underline{1.30} & 34.28 & \underline{36.31} \\
12 & 384 & 0.38 & 55.42 & 71.07 & 78.56 & 87.36 & 1.33 & 53.79 & 69.73 & 78.66 & 87.24 & 1.33 & \textbf{34.49} & 36.23 \\
12 & 512 & \textbf{0.35} & \underline{56.96} & \underline{72.62} & \underline{80.78} & \underline{88.21} & 1.30 & \textbf{55.74} & \underline{72.23} & \underline{79.91} & \underline{88.05} & \underline{1.30} & 33.93 & 36.23 \\
12 & 768 & 0.38 & 56.78 & 71.78 & 80.01 & 88.01 & \underline{1.29} & 53.71 & 70.70 & 79.46 & 87.89 & 1.34 & \underline{34.32} & \textbf{36.50} \\
16 & 512 & 0.39 & \textbf{57.51} & \textbf{73.17} & \textbf{81.19} & \textbf{88.88} & \textbf{1.26} & \underline{54.85} & \textbf{72.50} & \textbf{80.82} & \textbf{88.38} & \textbf{1.25} & 33.58 & 35.78 \\
\bottomrule
\end{tabular}
\caption{Model capacity ablation on the mixed test set (HumanML3D + \dbname, $N{=}5088$). $\downarrow$ = lower is better, $\uparrow$ = higher is better; $\rightarrow$ (Div/APD): closer to GT is better. Best in \textbf{bold}, second best \underline{underlined}.}
\label{tab:abl_size_mixed}
\end{table*}

\begin{table*}[!ht]
\centering
\small
\setlength{\tabcolsep}{5pt}
\begin{tabular}{@{}l l c c c c c c c c c c c cc@{}}
\toprule
\multirow{2}{*}{\textbf{Depth}} & \multirow{2}{*}{\textbf{Width}} & \multirow{2}{*}{\textbf{FID}$\downarrow$}
    & \multicolumn{5}{c}{\textbf{C2M}}
    & \multicolumn{5}{c}{\textbf{M2C}}
    & \multirow{2}{*}{\textbf{Div}$\rightarrow$} & \multirow{2}{*}{\textbf{APD}$\rightarrow$} \\
\cmidrule(lr){4-8}\cmidrule(lr){9-13}
    & & & \textbf{R@1}$\uparrow$ & \textbf{R@2}$\uparrow$ & \textbf{R@3}$\uparrow$ & \textbf{R@5}$\uparrow$ & \textbf{MedR}$\downarrow$
      & \textbf{R@1}$\uparrow$ & \textbf{R@2}$\uparrow$ & \textbf{R@3}$\uparrow$ & \textbf{R@5}$\uparrow$ & \textbf{MedR}$\downarrow$ & \\
\midrule
GT & -- & 0.01 & 40.18 & 59.64 & 70.54 & 84.64 & 2.00 & 40.00 & 63.84 & 76.16 & 87.32 & 1.80 & 39.69 & 40.41 \\
\midrule
8 & 512 & 0.76 & 36.96 & 56.52 & 69.11 & 83.48 & 2.11 & \underline{36.70} & 57.68 & 72.50 & \underline{86.61} & \underline{2.06} & 35.91 & 36.94 \\
12 & 384 & \underline{0.71} & 35.62 & 55.36 & 67.95 & 84.64 & 2.29 & 35.45 & 55.54 & 70.18 & 85.62 & 2.17 & \underline{36.48} & 37.25 \\
12 & 512 & 0.73 & \underline{37.05} & \underline{58.75} & \textbf{71.96} & \textbf{85.98} & 2.11 & \textbf{37.95} & \textbf{61.25} & \textbf{73.84} & \textbf{87.95} & \textbf{2.03} & 36.34 & \underline{37.48} \\
12 & 768 & \textbf{0.58} & 36.34 & 55.98 & 68.93 & 84.73 & \underline{2.03} & 34.29 & 56.96 & 71.52 & 85.80 & 2.23 & \textbf{38.62} & \textbf{39.02} \\
16 & 512 & 0.73 & \textbf{38.21} & \textbf{59.38} & \underline{71.34} & \underline{85.89} & \textbf{2.00} & \underline{36.70} & \underline{58.93} & \underline{72.77} & 86.52 & \textbf{2.03} & 36.38 & 37.35 \\
\bottomrule
\end{tabular}
\caption{Model capacity ablation on the \dbname test set ($N{=}1120$). $\downarrow$ = lower is better, $\uparrow$ = higher is better; $\rightarrow$ (Div/APD): closer to GT is better. Best in \textbf{bold}, second best \underline{underlined}.}
\label{tab:abl_size_stam}
\end{table*}

\begin{table*}[!ht]
\centering
\small
\setlength{\tabcolsep}{5pt}
\begin{tabular}{@{}l l c c c c c c c c c c c cc@{}}
\toprule
\multirow{2}{*}{\textbf{Depth}} & \multirow{2}{*}{\textbf{Width}} & \multirow{2}{*}{\textbf{FID}$\downarrow$}
    & \multicolumn{5}{c}{\textbf{C2M}}
    & \multicolumn{5}{c}{\textbf{M2C}}
    & \multirow{2}{*}{\textbf{Div}$\rightarrow$} & \multirow{2}{*}{\textbf{APD}$\rightarrow$} \\
\cmidrule(lr){4-8}\cmidrule(lr){9-13}
    & & & \textbf{R@1}$\uparrow$ & \textbf{R@2}$\uparrow$ & \textbf{R@3}$\uparrow$ & \textbf{R@5}$\uparrow$ & \textbf{MedR}$\downarrow$
      & \textbf{R@1}$\uparrow$ & \textbf{R@2}$\uparrow$ & \textbf{R@3}$\uparrow$ & \textbf{R@5}$\uparrow$ & \textbf{MedR}$\downarrow$ & \\
\midrule
GT & -- & 0.01 & 56.22 & 71.17 & 78.78 & 85.89 & 1.17 & 54.56 & 69.91 & 77.04 & 84.75 & 1.25 & 34.92 & 35.92 \\
\midrule
8 & 512 & \underline{0.45} & 60.76 & 74.24 & 81.20 & 87.95 & 1.10 & 59.58 & 74.34 & 81.05 & 87.30 & \underline{1.10} & \textbf{34.90} & \textbf{35.32} \\
12 & 384 & 0.48 & 60.99 & 75.25 & 81.25 & 87.83 & 1.11 & 58.77 & 73.51 & 80.32 & 87.22 & \underline{1.10} & 34.51 & \underline{35.12} \\
12 & 512 & \textbf{0.44} & \underline{62.42} & \underline{76.29} & \underline{82.84} & 88.53 & 1.10 & \textbf{60.43} & \underline{75.05} & 81.17 & 87.80 & \underline{1.10} & \underline{34.60} & 35.00 \\
12 & 768 & 0.49 & 62.30 & 76.06 & 82.76 & \underline{88.73} & \underline{1.06} & 58.97 & 74.40 & \underline{81.50} & \underline{88.23} & 1.12 & 34.15 & 34.96 \\
16 & 512 & 0.48 & \textbf{62.83} & \textbf{77.14} & \textbf{83.39} & \textbf{89.24} & \textbf{1.03} & \underline{59.93} & \textbf{75.60} & \textbf{82.36} & \textbf{88.28} & \textbf{1.05} & 33.90 & 34.46 \\
\bottomrule
\end{tabular}
\caption{Model capacity ablation on the HumanML3D test set ($N{=}3968$). $\downarrow$ = lower is better, $\uparrow$ = higher is better; $\rightarrow$ (Div/APD): closer to GT is better. Best in \textbf{bold}, second best \underline{underlined}.}
\label{tab:abl_size_hml}
\end{table*}

\subsection{Ablation on Model Capacity}

As shown in Tab.~\ref{tab:abl_size_mixed}, \ref{tab:abl_size_stam}, and \ref{tab:abl_size_hml}, reducing the model capacity does not yield consistent improvements and generally leads to comparable or degraded performance. Increasing the model size can improve individual metrics, but these gains do not consistently translate across evaluation criteria or test splits.

For example, comparing Rows~4 and~5, the larger $12\times768$ model achieves notably better FID, Diversity, and APD on the \dbname test set, but does not provide consistently better R-Precision or overall performance across the other evaluation splits compared with the $12\times512$ baseline. Similarly, comparing Rows~4 and~6, the deeper $16\times512$ model generally achieves slightly higher R-Precision, but this improvement is accompanied by a modest degradation in FID. These results suggest that simply increasing model capacity does not provide a consistent benefit, and we therefore adopt the $12\times512$ configuration as a favorable balance between model capacity and overall performance.

\begin{table*}[!ht]
\centering
\small
\setlength{\tabcolsep}{5pt}
\begin{tabular}{@{}l c c c c c c c c c c c cc@{}}
\toprule
\multirow{2}{*}{\textbf{CFG}} & \multirow{2}{*}{\textbf{FID}$\downarrow$}
    & \multicolumn{5}{c}{\textbf{C2M}}
    & \multicolumn{5}{c}{\textbf{M2C}}
    & \multirow{2}{*}{\textbf{Div}$\rightarrow$} & \multirow{2}{*}{\textbf{APD}$\rightarrow$} \\
\cmidrule(lr){3-7}\cmidrule(lr){8-12}
    & & \textbf{R@1}$\uparrow$ & \textbf{R@2}$\uparrow$ & \textbf{R@3}$\uparrow$ & \textbf{R@5}$\uparrow$ & \textbf{MedR}$\downarrow$
      & \textbf{R@1}$\uparrow$ & \textbf{R@2}$\uparrow$ & \textbf{R@3}$\uparrow$ & \textbf{R@5}$\uparrow$ & \textbf{MedR}$\downarrow$ & \\
\midrule
GT & 0.01 & 52.79 & 68.69 & 77.14 & 85.89 & 1.33 & 51.47 & 68.69 & 77.14 & 85.36 & 1.38 & 35.84 & 37.51 \\
\midrule
1.0 & 0.44 & 52.38 & 68.30 & 76.65 & 85.20 & 1.37 & 51.47 & 67.63 & 76.02 & 85.57 & 1.39 & 32.54 & 34.57 \\
1.5 & 0.37 & 55.96 & 71.48 & 79.52 & 87.62 & 1.31 & 54.78 & 70.77 & 78.87 & 87.60 & 1.31 & 33.39 & 35.66 \\
2.0 & \textbf{0.35} & 56.96 & 72.62 & 80.78 & 88.21 & \underline{1.30} & 55.78 & 72.21 & 79.89 & 88.07 & 1.30 & 33.93 & 36.23 \\
2.5 & \textbf{0.35} & \textbf{57.59} & \underline{73.45} & 81.47 & 88.58 & 1.31 & 56.45 & 73.05 & 80.29 & 88.76 & \textbf{1.26} & 34.20 & 36.54 \\
3.0 & \underline{0.36} & \underline{57.57} & \textbf{73.78} & \textbf{81.80} & \textbf{88.72} & \underline{1.30} & \textbf{56.86} & \textbf{73.29} & \underline{80.54} & \textbf{88.99} & \underline{1.28} & \underline{34.48} & \underline{36.77} \\
3.5 & 0.37 & 57.53 & \underline{73.45} & \underline{81.60} & \underline{88.68} & \textbf{1.28} & \underline{56.66} & \underline{73.09} & \textbf{80.78} & \underline{88.84} & \underline{1.28} & \textbf{34.66} & \textbf{36.94} \\
\bottomrule
\end{tabular}
\caption{CFG-scale sweep on the mixed test set (HumanML3D + \dbname, $N{=}5088$). $\downarrow$ = lower is better, $\uparrow$ = higher is better; $\rightarrow$ (Div/APD): closer to GT is better. Best in \textbf{bold}, second best \underline{underlined}.}
\label{tab:cfg_mixed}
\end{table*}

\begin{table*}[!ht]
\centering
\small
\setlength{\tabcolsep}{5pt}
\begin{tabular}{@{}l c c c c c c c c c c c cc@{}}
\toprule
\multirow{2}{*}{\textbf{CFG}} & \multirow{2}{*}{\textbf{FID}$\downarrow$}
    & \multicolumn{5}{c}{\textbf{C2M}}
    & \multicolumn{5}{c}{\textbf{M2C}}
    & \multirow{2}{*}{\textbf{Div}$\rightarrow$} & \multirow{2}{*}{\textbf{APD}$\rightarrow$} \\
\cmidrule(lr){3-7}\cmidrule(lr){8-12}
    & & \textbf{R@1}$\uparrow$ & \textbf{R@2}$\uparrow$ & \textbf{R@3}$\uparrow$ & \textbf{R@5}$\uparrow$ & \textbf{MedR}$\downarrow$
      & \textbf{R@1}$\uparrow$ & \textbf{R@2}$\uparrow$ & \textbf{R@3}$\uparrow$ & \textbf{R@5}$\uparrow$ & \textbf{MedR}$\downarrow$ & \\
\midrule
GT & 0.01 & 40.18 & 59.64 & 70.54 & 84.64 & 2.00 & 40.00 & 63.84 & 76.16 & 87.32 & 1.80 & 39.69 & 40.41 \\
\midrule
1.0 & 0.83 & 36.96 & 57.14 & 71.16 & 85.54 & \textbf{2.06} & 36.70 & 59.55 & 72.59 & 87.41 & \underline{2.00} & 35.69 & 36.90 \\
1.5 & \underline{0.74} & \textbf{37.14} & 58.30 & 71.88 & \underline{86.70} & \underline{2.09} & 37.05 & \underline{60.98} & \underline{73.57} & \textbf{87.86} & 2.09 & 36.24 & 37.42 \\
2.0 & \textbf{0.73} & \underline{37.05} & 58.75 & \underline{71.96} & 85.98 & 2.11 & 37.95 & \textbf{61.25} & \textbf{73.75} & \textbf{87.86} & 2.03 & \textbf{36.34} & \textbf{37.48} \\
2.5 & \textbf{0.73} & 36.88 & 58.84 & \textbf{72.23} & 85.98 & 2.14 & \underline{38.04} & 60.89 & 72.95 & 87.59 & \textbf{1.97} & \textbf{36.34} & \underline{37.43} \\
3.0 & \underline{0.74} & 36.79 & \underline{59.11} & 71.25 & 86.43 & 2.20 & 37.95 & 60.27 & 72.59 & \underline{87.68} & 2.06 & \underline{36.33} & 37.39 \\
3.5 & \underline{0.74} & 36.43 & \textbf{59.20} & 70.98 & \textbf{86.79} & 2.14 & \textbf{38.12} & 60.45 & 72.41 & \underline{87.68} & 2.06 & 36.32 & 37.35 \\
\bottomrule
\end{tabular}
\caption{CFG-scale sweep on the \dbname test set ($N{=}1120$). $\downarrow$ = lower is better, $\uparrow$ = higher is better; $\rightarrow$ (Div/APD): closer to GT is better. Best in \textbf{bold}, second best \underline{underlined}.}
\label{tab:cfg_stam}
\end{table*}

\begin{table*}[!ht]
\centering
\small
\setlength{\tabcolsep}{5pt}
\begin{tabular}{@{}l c c c c c c c c c c c cc@{}}
\toprule
\multirow{2}{*}{\textbf{CFG}} & \multirow{2}{*}{\textbf{FID}$\downarrow$}
    & \multicolumn{5}{c}{\textbf{C2M}}
    & \multicolumn{5}{c}{\textbf{M2C}}
    & \multirow{2}{*}{\textbf{Div}$\rightarrow$} & \multirow{2}{*}{\textbf{APD}$\rightarrow$} \\
\cmidrule(lr){3-7}\cmidrule(lr){8-12}
    & & \textbf{R@1}$\uparrow$ & \textbf{R@2}$\uparrow$ & \textbf{R@3}$\uparrow$ & \textbf{R@5}$\uparrow$ & \textbf{MedR}$\downarrow$
      & \textbf{R@1}$\uparrow$ & \textbf{R@2}$\uparrow$ & \textbf{R@3}$\uparrow$ & \textbf{R@5}$\uparrow$ & \textbf{MedR}$\downarrow$ & \\
\midrule
GT & 0.01 & 56.22 & 71.17 & 78.78 & 85.89 & 1.17 & 54.56 & 69.91 & 77.04 & 84.75 & 1.25 & 34.92 & 35.92 \\
\midrule
1.0 & 0.54 & 56.53 & 71.19 & 77.77 & 84.68 & 1.21 & 55.29 & 69.43 & 76.66 & 84.78 & 1.23 & 32.65 & 32.99 \\
1.5 & 0.46 & 61.11 & 74.85 & 81.20 & 87.60 & 1.11 & 59.35 & 73.34 & 79.81 & 87.27 & 1.11 & 33.88 & 34.28 \\
2.0 & \textbf{0.44} & 62.42 & 76.29 & 82.84 & 88.53 & 1.10 & 60.48 & 75.03 & 81.17 & 87.83 & 1.10 & \underline{34.60} & 35.00 \\
2.5 & \textbf{0.44} & \textbf{63.23} & 77.27 & 83.62 & 88.99 & 1.09 & 61.44 & 76.13 & 81.80 & 88.79 & \textbf{1.07} & \textbf{35.04} & 35.43 \\
3.0 & \underline{0.45} & \underline{63.21} & \textbf{77.77} & \textbf{84.22} & \textbf{89.21} & \underline{1.08} & \textbf{61.97} & \textbf{76.61} & \underline{82.26} & \textbf{89.14} & \underline{1.09} & 35.34 & \underline{35.74} \\
3.5 & 0.47 & \underline{63.21} & \underline{77.44} & \underline{83.92} & \underline{89.11} & \textbf{1.06} & \underline{61.74} & \underline{76.26} & \textbf{82.56} & \underline{88.99} & 1.10 & 35.57 & \textbf{35.98} \\
\bottomrule
\end{tabular}
\caption{CFG-scale sweep on the HumanML3D test set ($N{=}3968$). $\downarrow$ = lower is better, $\uparrow$ = higher is better; $\rightarrow$ (Div/APD): closer to GT is better. Best in \textbf{bold}, second best \underline{underlined}.}
\label{tab:cfg_hml}
\end{table*}

\subsection{CFG Scale}

As shown in Tab.~\ref{tab:cfg_mixed}, \ref{tab:cfg_stam}, and \ref{tab:cfg_hml}, we evaluate CFG scales from $1.0$ to $3.5$ at intervals of $0.5$. Overall, CFG scales between $2.0$ and $3.0$ provide the most favorable balance across the evaluation metrics, with CFG $=2.0$ achieving particularly strong FID. Increasing the CFG scale beyond $3.0$ can yield slightly higher R-Precision, Diversity, or APD, but these improvements are generally accompanied by a modest degradation in FID. We therefore adopt CFG $=2.0$ as the default inference setting.

\end{document}